\documentclass[12pt]{article}
\usepackage{graphicx}      
\usepackage{setspace}      
\usepackage{enumerate}     
\usepackage{amssymb,amsthm,amsmath,amsfonts,amsbsy}
\usepackage{mathrsfs,syntonly,graphicx,color}
\usepackage{setspace,geometry}
\usepackage{indentfirst}
\usepackage{bbm}
\usepackage{bm}
\usepackage{float}
\usepackage{tabularx}      
\usepackage{tikz}
\usepackage{ctable}        
\usepackage{multirow}      
\usepackage{comment}       
\usepackage[round]{natbib}
\usepackage{caption}
\usepackage{algorithm}    
\usepackage{algpseudocode}  
\usepackage[colorlinks, urlcolor=blue, linkcolor=blue,citecolor=blue]{hyperref} 
\usepackage{subcaption} 
\usepackage{soul} 
\usepackage{lineno}
\usepackage{lscape} 
\usepackage{chngcntr} 
\usepackage[labelfont=bf]{caption} 
\usepackage[flushleft]{threeparttable} 
\usepackage{authblk}
\usepackage{diagbox}
\usepackage{dcolumn}

\usepackage{multibib}
\newcites{app}{Appendix References}

\title{The Anatomy of a Blockchain Prediction Market: Polymarket in the 2024 U.S. Presidential Election\thanks{This paper was previously circulated under the title ``The Anatomy of Polymarket: Evidence from the 2024 Presidential Election''.}}
\author{Kwok Ping Tsang and Zichao Yang\thanks{Tsang: Department of Economics, Virginia Tech,
		Pamplin Hall, Blacksburg, VA, 24061, USA; E-mail: byront@vt.edu. Yang: Wenlan School of Business, Zhongnan University of
		Economics and Law, Wuhan, China; E-mail: yang\_zichao@outlook.com.}}
\date{\today}

\begin{document}

\maketitle

\begin{abstract}

\noindent Using the complete on-chain settlement ledger, we study Polymarket's 2024 U.S. presidential election markets, focusing on the Trump YES price widely quoted as a forecast of the race. Because the platform mints and burns outcome shares inside ordinary trades, naive aggregation overstates turnover. We develop a transaction-level decomposition that separates turnover from net inflow and market activity: in October it reports \$391 million of Trump-market turnover against \$958 million of naive volume. Corrected price impact reveals a shallower market: moving the October forecast by five percentage points cost \$9.1 million rather than \$15.6 million. Capital entered on both sides of the race throughout the month, consistent with heterogeneous beliefs rather than one-sided manipulation. The overstatement holds across 249 markets and is largest in thin, young ones.

~

\noindent \textbf{JEL Codes}: D47; G12; G14; D83.

\noindent \textbf{Keywords}: Prediction markets; trading volume; market microstructure; market manipulation; price impact.
\end{abstract}
\clearpage
\onehalfspacing
\section{Introduction}
\label{sec:intro}

\noindent Polymarket's event page for the 2024 U.S. presidential election displayed \$3.69 billion of volume.\footnote{See \href{https://polymarket.com/event/presidential-election-winner-2024}{Polymarket: Presidential Election Winner 2024}, which displays \$3,686,335,059 for the event.} Figures like this were widely read as evidence that the market was too deep for any single trader to move, and its implied probabilities were quoted by major news outlets as a real-time forecast of the race. In October 2024, news reports revealed that a small set of accounts had placed tens of millions of dollars in pro-Trump bets. The public debate turned on whether those bets were distorting the price that voters, donors, and journalists read as the probability of a Trump victory.\footnote{1. \href{https://www.wsj.com/finance/betting-election-pro-trump-ad74aa71?mod=article_inline}{Wall Street Journal: A Mystery \$30 Million Wave of Pro-Trump Bets Has Moved a Popular Prediction Market}. 2. \href{https://www.reuters.com/world/polymarket-says-mystery-trump-bettor-is-french-national-2024-10-24/}{Reuters: Polymarket says mystery Trump bettor is French national}. 3. \href{https://www.cnbc.com/2024/10/24/polymarket-trump-french-election-bet.html}{CNBC: French trader bet over \$28 million on Trump election win using 4 Polymarket accounts}.} Both sides of that debate rest on a measurement question: how much trading actually stands behind the price. We show that aggregating the ledger's dollar flows overstates the Trump market's exchange-equivalent turnover by roughly a factor of two and a half. Correcting the overstatement shows a market easier to move than its volume implied.

The overstatement is built into how Polymarket settles trades. Outcome shares are tokens that the settlement layer mints and burns on demand inside ordinary matched transactions: a YES buyer meeting a NO buyer creates a new pair, and a YES seller meeting a NO seller destroys one. The supply of traded claims is therefore endogenous, and raw transaction flows mix secondary-market exchange with primary issuance and redemption. An analyst reading the raw ledger faces the same problem that \citet{andersondyl2007} document for equity markets. There, dealer-market conventions led NASDAQ to report roughly three times the turnover of comparable NYSE stocks, and practitioners applied ad hoc adjustment factors for decades \citep{huangstoll1996}. In both settings the reported number feeds real decisions. Commentators used it to judge whether the price could be trusted, and a growing academic literature downloads the same on-chain records and treats aggregated flows as turnover.

The paper's contribution starts before any volume figure, with how to read the ledger. Every match Polymarket settles leaves a public record, and we show that the record reveals each transaction's route: whether it exchanged existing shares, minted new ones, or burned old ones. The route is what summing the records ignores. An exchanged dollar enters the ledger twice, once on each side of the match, while a minted or burned dollar enters once. Summing the raw records therefore double counts exchange and conflates it with primary flows, a concern \citet{slivkoff2025paradigm} has also raised. \autoref{subsec:trade_mech_hetero} works through the mechanics with examples from the ledger.

The remedy is a transaction-level decomposition. Every matched transaction splits into an exchange component, a minting component, and a burning component. Aggregating the components yields three market-level measures that answer different economic questions. Exchange-equivalent turnover $V^E$ captures how intensively existing claims are reshuffled among traders, net inflow $F$ how much new risk-bearing capital enters or leaves, and market activity $V^G$ the two combined. An exact accounting identity ties the measures to the naive dollar aggregate: at any horizon, it equals turnover plus market activity. October 2024 makes it concrete: \$958 million of naive volume in the Trump market is exactly \$391 million of turnover plus \$567 million of market activity.

With the flows correctly measured, we ask what the naive volume figure obscured: how costly the Trump price was to move, and whether concentrated buying alone drove the forecast. We take the cost first, because it sets the scale on which the concentration of buying is judged. The market was cheaper to move than its volume implied, throughout the period in which its price was quoted as the election forecast. We estimate Kyle's $\lambda$ \citep{kyle1985continuous} for the Trump YES share month by month, once on the naive flow and once on the corrected flow, holding prices and trade signs identical so that the gap between the two estimates isolates the flow measure. The corrected $\hat\lambda$ exceeds the naive one in every month, by a factor of 1.1 to 2.1. The gap matters most where the market was thinnest: in the first half of 2024 a \$1 million net purchase moved the implied probability by 4 to 15 percentage points, against about half a point in October. Moving the October forecast by five percentage points cost \$9.1 million, not \$15.6 million. Across markets, a dollar of secondary-market trading moves the price more than a dollar of primary-market flow.

Second, concentrated buying did not drive the forecast alone. Although the reported whale's October purchases equaled about 15 percent of the Trump market's net inflow, capital kept entering on the Democratic side as those positions accumulated, and participation stayed broad and two-sided.

The depth and participation findings center on the Trump market, which raises the question of whether the measurement error behind them is a quirk of that market. It is not. The measurement error is systematic across Polymarket's election markets and varies with market maturity. Applying the decomposition to all 249 markets, we find that the median active market, one with at least \$0.1 million of naive volume over its life, records three times as much naive volume as exchange-equivalent turnover. The factor ranges from about two to fifty across the election-related sample. This variation rules out a fixed multiplier across markets. The overstatement is largest in smaller and younger markets, where issuance is high relative to turnover, so naive volume is most misleading where secondary trading is least developed.

The paper speaks to three literatures. The prediction-market literature has long studied prices as belief aggregators \citep{wolfers2004prediction, berg2008accuracy, arrow2008promise} and the conditions under which prices map to probabilities \citep{manski2006interpreting, wolfers2006interpreting}, with transaction-level work showing that microstructure matters for reading them \citep{rothschild2015trading}. That literature treats volume as a given object. In tokenized markets, volume has to be constructed before liquidity, disagreement, or manipulation can be studied. The market-manipulation literature analyzes when trades can move event-market prices and to what effect \citep{allen1992stock, hansen2004manipulation, hanson2006manipulation, rothschild2015trading}, and our price-impact application quantifies the capital cost of doing so in the most prominent political market to date. Emerging research on decentralized prediction-market design documents how trading and settlement architecture vary across venues \citep{rahman2026sok}. Recent empirical studies use Polymarket to analyze price discovery, political motives, and wash trading \citep{ng2026price, chen2024political, sirolly2025network}. These questions require activity measures that distinguish exchange from issuance. Our framework supplies that measurement layer, and the applications demonstrate that using it changes conclusions.

The remainder of the paper proceeds as follows. \autoref{polymarket} describes Polymarket's market design and trading mechanisms. \autoref{sec:volume_measurement} shows why the naive aggregates fail and constructs the corrected measures. \autoref{sec:anatomy} reads the Presidential Election Winner event through both the corrected measures and the naive aggregates, and establishes the wedge's scope across 249 election markets. \autoref{results_lambda} measures market depth and price formation: the cost of moving the forecast, and which flow carries more price impact per dollar. \autoref{sec:traders} turns to participation as the market deepened. \autoref{conclusion} concludes.

\section{Polymarket: Market Design and Trade Mechanisms}
\label{polymarket}

\subsection{Platform architecture and event formats}

\noindent Polymarket is a hybrid-decentralized prediction-market platform launched in 2020. The platform combines an off-chain order book with on-chain settlement. The design has two primary components:
\begin{enumerate}
	\item Off-chain central limit order book (CLOB): A centralized ``operator'' manages the order book and matches buy and sell orders off-chain. Traders can submit and cancel orders without paying blockchain transaction fees for each order-management action.
	\item On-chain settlement layer: Smart contracts deployed on Polygon settle matched trades, hold collateral, and manage outcome positions. During our sample, collateral was denominated in USDC, a stablecoin pegged one-for-one to the U.S. dollar.
\end{enumerate}

Following Polymarket's convention, we use \emph{event} for the top-level prediction topic and \emph{market} for each tradable binary question within it. Every market has a YES share and a NO share. The active central-limit-order-book events relevant to this paper have two structures and use two corresponding settlement formats:\footnote{Legacy scalar markets are not part of our sample. They were typically automated-market-maker markets and were no longer active during the period we study.}
\begin{enumerate}
	\item Single-market event: An event containing one market. Its YES and NO shares represent the two possible answers to the market's binary question, and the winning share pays 1 USDC at resolution. Trades settle through the \texttt{CTFExchange} contract.
	\item Multi-market event: An event containing two or more related binary markets. In the multi-market events studied here, the underlying outcomes are mutually exclusive, exactly one market resolves YES, and that market's YES share pays 1 USDC, and each other market's NO share pays 1 USDC. Polymarket calls a set of linked mutually exclusive markets a negative-risk group, which gives the exchange contract that settles their trades its name, \texttt{NegRiskCTFExchange}.\footnote{Polymarket's user-facing documentation also classifies markets by outcome count, calling a single-market event a binary market and a multi-market event a categorical market. The second label applies the word ``market'' to an entire event. We follow the platform's data model instead, in which each tradable binary question is a separate market. This convention matches how the settlement records we analyze are organized, and it keeps the three levels distinct: an event contains markets, and each market issues YES and NO shares.}
\end{enumerate}

The \href{https://polymarket.com/event/presidential-election-winner-2024}{2024 Presidential Election Winner} event is a multi-market event and contains 17 candidate markets. Trump, Biden, and Harris markets account for the vast majority of activity in this event, so the main analysis follows these three markets. Each market's YES share pays 1 USDC if that candidate wins and zero otherwise, while its NO share pays 1 USDC if the candidate does not win. Before resolution, both shares can be traded. A YES-NO pair is therefore a complete set of claims backed by 1 USDC of collateral.

\subsection{Order matching and position mechanics}

Every order on Polymarket is treated as a limit order, filled at its stated price or better. The operator matches these orders off-chain, and one taker order may be filled against multiple maker orders. The relevant exchange contract then settles the resulting match atomically on Polygon. Each matched order produces an \texttt{OrderFilled} log, and the full match produces an \texttt{OrdersMatched} summary log.

Once orders are matched, the settlement contract can fill them through three economic routes: (1) when a buyer of YES meets a seller of YES, an existing share changes hands through ordinary secondary-market exchange. The same applies to a buyer and seller of NO; (2) when a buyer of YES meets a buyer of NO and their payments sum to 1 USDC, the contract locks the collateral, creates a new YES-NO pair, and distributes the two shares to the buyers. This route is primary issuance, which we call share minting; (3) when a seller of YES meets a seller of NO, the contract destroys their complete pair and releases the 1 USDC of collateral. This route is redemption, which we call share burning.\footnote{Polymarket calls the corresponding matched-order routes \emph{minting} and \emph{merging}. We use \emph{minting} and \emph{burning}, the standard terms in the crypto community for increasing and decreasing token supply.} A single matched transaction can combine exchange with minting or burning when one order is filled against several counterparty orders.

Traders choose the share, quantity, and price of their orders, but they do not choose the settlement route. The matching engine determines whether the orders are filled by transferring existing shares or by creating or destroying a complete pair. The same minting and burning operations are also available directly to arbitrageurs. If a YES-NO pair trades above 1 USDC, an arbitrageur can create a pair and sell both shares. If it trades below 1 USDC, an arbitrageur can buy both shares, destroy the pair, and redeem the collateral. These operations support the pricing identity $P^{\text{YES}} + P^{\text{NO}} = 1$ up to trading frictions.

Multi-market events add one further operation, conversion. A trader can convert equal quantities of NO shares across several markets into YES shares in the remaining markets plus USDC, which frees redundant collateral and supports the arbitrage that keeps prices consistent across the linked markets. For further technical detail on Polymarket's market structure and order-matching mechanism, see \autoref{appendix_protocol}.

\subsection{Data}

Polymarket's exchange contracts emit a public log for each filled order and matched transaction, and the full history can be read directly from the Polygon blockchain. We build the sample from these logs collected through the public \href{https://polygon-rpc.com/}{Polygon RPC} endpoint\footnote{Polymarket uses the \texttt{CTFExchange} contract for single-market events and the \texttt{NegRiskCTFExchange} contract for multi-market events.}. Each record identifies the block and the transaction, the order's maker wallet and its taker field, which is a counterparty wallet or the exchange contract, the asset identifiers of the two legs, and the amounts exchanged in shares and in USDC.\footnote{We retrieve wallet addresses from the logs, not the entities that control them. However, Polymarket accounts are created through a hosted interface and carry a display name, so splitting activity across many wallets is far less common here than on Bitcoin, where address reuse is discouraged and a single participant routinely controls many addresses. \citet{tsangyang2026political} search the same ledger for addresses potentially controlled by the same entity and find only 355 such addresses.}

Our main analysis uses the complete \texttt{OrderFilled} and \texttt{OrdersMatched} records for the Trump, Biden, and Harris markets from January 5, 2024, when these markets opened to trading, through November 6, 2024 at 06:46 UTC, when Fox News first called the race for Trump. The Trump market alone contains 2,348,771 matched transactions, which emitted 5,085,439 \texttt{OrderFilled} logs, one for each order filled, and 2,348,771 \texttt{OrdersMatched} logs, one for each transaction. For the cross-market analysis in \autoref{subsec:crossmarket}, we extend the sample to all 249 markets linked to the 2024 election.

\section{From Settlement Records to Market-Level Measures}
\label{sec:volume_measurement}

\noindent On Polymarket, exchange, minting, and burning settle through the same matched transactions and appear in the same logs. Those shared logs raise two measurement questions: how many dollars transfer existing risk between traders, and how many create or withdraw risk-bearing claims. A dollar on the ledger may be turnover of an existing claim, primary issuance, or redemption, so a naive aggregation of the raw records mixes the three activities together and answers neither question. A usable measure must first recover the route each dollar took. This section reads the routes off a single transaction, shows why both naive aggregates mismeasure them, and builds the measures that answer the two questions.

\subsection{Reading a matched transaction}
\label{subsec:trade_mech_hetero}

\autoref{polymarket} defines the three routes through which matched orders settle. The ledger, however, does not record the route. It records orders, and each route leaves its own signature across the orders of a single transaction. \autoref{tab_example_orders_types} shows examples of filled orders from the 2024 Presidential Election Winner event.

In each order, one asset is a YES or NO share of a certain market, identified by its unique token ID, and the other asset is USDC (asset ID \texttt{0}). The \emph{Simple Trade} consists of three matched orders. Traders \texttt{0x869} and \texttt{0xd42} sold 200 and 10 shares of Trump NO for 118 USDC and 5.9 USDC, respectively. The counterparty, trader \texttt{0x9d8}, bought 210 shares of Trump NO for 123.9 USDC. These orders were matched by the Polymarket \texttt{NegRiskCTFExchange} smart contract (\texttt{0xC5d}). All three orders involve the same share, and the 123.9 USDC paid on the buy side equals the amount received on the sell side.

In the \emph{Share Minting Trade}, trader \texttt{0x351} bought 6000 shares of Biden NO for 3960 USDC, and trader \texttt{0xE0D} bought 6000 shares of Biden YES for 2040 USDC. The two traders are buying the complementary YES and NO shares in the Biden market, so their combined 6000 USDC is locked as collateral and 6000 new pairs of Biden YES and NO shares are minted and distributed to them.

In the \emph{Share Burning Trade}, trader \texttt{0x64C} sold 206.19 shares of Trump NO for 123.714 USDC, and trader \texttt{0xff6} sold 206.19 shares of Trump YES for 82.476 USDC. Both traders sell and neither buys, so the 206.19 complete pairs are destroyed and the 206.19 USDC of collateral is released to the two sellers at their traded prices.

The \emph{Mixed Trade} example combines share exchange and share minting. Trader \texttt{0xf0b} bought 238.095237 shares of Trump YES for 99.999999 USDC. To fill this demand, the Polymarket \texttt{NegRiskCTFExchange} smart contract matched it with two orders: (1) trader \texttt{0x869} sold 200 shares of Trump YES for 84 USDC, and (2) trader \texttt{0xd42} bought 38.095237 shares of Trump NO for 22.095238 USDC. Hence, 200 shares of Trump YES changed hands through direct exchange, and 38.095237 new shares of Trump YES and NO were minted to fill the remaining demand. A mixed trade can also combine share exchange with share burning. Minting and burning of the same share, however, cannot occur in the same transaction.

Polymarket also reports an \texttt{OrdersMatched} summary of each transaction. \autoref{tab_example_trade_data} reports selected fields from the logs of the mixed trade. The \texttt{OrdersMatched} record identifies the initiating wallet but omits the resting-order participants and the full set of assets recorded in \texttt{OrderFilled}.

Based on the information reported in \texttt{OrderFilled} and \texttt{OrdersMatched}, there are two naive ways of calculating market volume on Polymarket. A first naive approach measures volume from the \texttt{OrdersMatched} logs alone. In the mixed-trade example, it assigns 99.999999 USDC of volume to Trump YES. That measure captures the buyer's total payment but omits the Trump NO volume created through minting, and it records only one side of pure minting and burning transactions. A second naive approach aggregates the \texttt{OrderFilled} logs. In the same example, it implies Trump YES volume of $84 + 99.999999 = 183.999999$ USDC, recovering the omitted NO leg but counting the exchange component twice, once through the seller's order and once through the buyer's order. The same double counting occurs in every simple trade. The two errors therefore vary with route composition: \texttt{OrdersMatched} misses minting and burning legs, while \texttt{OrderFilled} double-counts exchange.\footnote{Polymarket's own displayed volume applies the \texttt{OrdersMatched} logic in shares rather than dollars, accumulating the taker-order share quantity of each transaction. It is a market-level contract count, not comparable with the dollar measures of this paper.}

\subsection{Transaction-level volume decomposition}
\label{subsec:decomposition}

Both naive aggregates fail for the same reason: they sum orders without asking which route each dollar took. The remedy is to classify the route first, transaction by transaction. To compute the volume of a given transaction, we use the \texttt{OrderFilled} logs associated with that transaction. Suppose there are $I$ buy-side orders (i.e., $\texttt{makerAssetId}_i=0$) and $J$ sell-side orders (i.e., $\texttt{takerAssetId}_j=0$). We define two gross USDC flows\footnote{Raw amounts are denominated in $10^{-6}$ USDC and we rescale them to USDC here.}:
\begin{align}
\text{buy vol} &=\frac{1}{10^6}\sum_{i=1}^{I}\texttt{makerAmountFilled}_i,\\
\text{sell vol} &=\frac{1}{10^6}\sum_{j=1}^{J}\texttt{takerAmountFilled}_j.
\end{align}

If $\text{buy vol} = \text{sell vol}$, then the transaction is a pure share exchange (as in the \emph{Simple Trade} example in \autoref{tab_example_orders_types}). In this case, only one type of share changes hands and the exchange component is:
\begin{equation}
\text{trade vol}=\min\{\text{buy vol},\text{sell vol}\}.
\end{equation}

When $\text{buy vol}>\text{sell vol}$, the transaction contains minting activity: it is either a \emph{Share Minting} transaction (with $\text{sell vol}=0$) or a \emph{Mixed Trade} (with $\text{sell vol}>0$ but $\text{sell vol}<\text{buy vol}$), as shown in \autoref{tab_example_orders_types}. We distinguish these cases using the number of distinct asset identifiers on the maker side: if \texttt{makerAssetId} takes more than one unique value, we classify the transaction as a \emph{Mixed Trade}. Otherwise it is a \emph{Share Minting}. In both cases, the market has two complementary outcome shares, denoted $(X_Y,X_N)$, and we compute exchange and minting components separately:

(1) If the maker-side asset identifier takes only one value (i.e., \texttt{0}, corresponding to USDC), then the transaction is a \emph{Share Minting}, and the exchange component is:
\begin{equation}
\text{trade vol} = \min\{\text{buy vol},\text{sell vol}\} = 0.
\end{equation}
The minting components for $(X_Y,X_N)$ are then:
\begin{align}
		\text{YES mint vol} &=\frac{1}{10^6}\sum_{i\in X_Y}\texttt{makerAmountFilled}_i,\\
		\text{NO mint vol} &=\frac{1}{10^6}\sum_{i\in X_N}\texttt{makerAmountFilled}_i.
\end{align}

(2) If \texttt{makerAssetId} takes multiple unique values, then the transaction is a \emph{Mixed Trade}, and the exchange component is:
\begin{equation}
\text{trade vol} = \min\{\text{buy vol},\text{sell vol}\}.
\end{equation}
Let the non-$\texttt{0}$ token identifier on the maker side be $X_a$, with $a\in\{Y,N\}$, and let $X_{\bar a}$ denote the complementary share. The corresponding minting components are:
\begin{align}
		\text{mint vol}(X_a) &=\frac{1}{10^6}\sum_{i\in X_a}\texttt{makerAmountFilled}_i - \text{trade vol},\\
		\text{mint vol}(X_{\bar a}) &=\frac{1}{10^6}\sum_{i\in X_{\bar a}}\texttt{makerAmountFilled}_i.
\end{align}

When $\text{buy vol}<\text{sell vol}$, the transaction may correspond to a \emph{Share Burning} or a \emph{Mixed Trade}, and the decomposition is analogous, yielding exchange and burning components for $(X_Y,X_N)$. Overall, we decompose each transaction into six components: $\texttt{YES trade vol}$, $\texttt{NO trade vol}$, $\texttt{YES mint vol}$, $\texttt{NO mint vol}$, $\texttt{YES burn vol}$, and $\texttt{NO burn vol}$. \autoref{appendix_tx_decomposition} shows the pseudocode of the transaction-level volume decomposition.

\subsection{Market-level measures}
\label{subsec:market_measures}

Aggregating the transaction-level components yields three market-level measures. Exchange-equivalent turnover answers the first question posed at the start of this section: how many dollars transferred existing risk between traders. Net inflow answers the second: how much risk-bearing capital entered or left the market on net. Gross market activity combines the two margins into a total that avoids double counting. We define each in turn. The exchange-equivalent turnover ($V^E$) for the Trump YES shares over an interval $T$ is:
\begin{align}
		V_{YES}^E(T) = & \sum_{\tau\in T}\text{YES trade vol}_{\tau}  \notag \\
		& + \min\Bigl(\sum_{\tau\in T}\text{YES mint vol}_{\tau}, \sum_{\tau\in T}\text{YES burn vol}_{\tau}\Bigr).
\end{align}
Exchange-equivalent turnover treats offsetting mint and burn flows over the interval as secondary-market trading routed through the settlement contract, counting each such transfer once. Because Polymarket's matching algorithm can determine whether a given order is filled through direct exchange or share minting, the split between trade and mint volumes partly reflects the matching mechanism rather than trader intent. Matching aggregate mints against aggregate burns is therefore a reasonable approximation of counterfactual secondary-market volume. \autoref{appendix_uniqueness} shows that, under the conditions stated there, exchange-equivalent turnover is the only window measure that counts each transferred dollar exactly once.

The net inflow ($F$) for the Trump YES shares over an interval $T$ is:
\begin{align}
		F_{YES}(T) = \sum_{\tau\in T}\text{YES mint vol}_{\tau} - \sum_{\tau\in T}\text{YES burn vol}_{\tau}.
\end{align}

A positive value of $F_{\text{YES}}$ indicates fresh capital committed on net by YES-side traders through share minting. A negative value indicates capital withdrawn through net burning, as traders redeem positions for collateral.

The gross market activity ($V^G$) for the Trump YES shares over an interval $T$ is:
\begin{align}
	V_{YES}^G(T)  = & \sum_{\tau\in T}\text{YES trade vol}_{\tau} \notag \\
	& \quad + \max\Bigl(\sum_{\tau\in T}\text{YES mint vol}_{\tau}, \sum_{\tau\in T}\text{YES burn vol}_{\tau}\Bigr) \notag \\
	= & V_{YES}^E(T)  + |F_{YES}(T)|.
\end{align}
While $V^E$ captures the reshuffling of existing positions among traders, $V^G$ also incorporates the net flow of new capital into or out of the market. Each definition is stated for the YES shares and applies unchanged to the NO shares. 

A market's measures are the sums of the measures for its two share types:
\begin{equation}
X(T) = X_{YES}(T) + X_{NO}(T), \qquad X \in \{V^E, F, V^G\}.
\end{equation}
Net inflow adds the two share types with their signs, because it measures the collateral the market gained or lost on net. Market activity adds their unoffset issuance without signs. Capital committed through one share type and withdrawn through the other are both activity, and a measure of how much business the market did should not let them offset. Figures reported below are market-level unless a share type is named.

All three measures are defined over an interval, but its length affects only turnover and market activity. Signed net inflow is additive, so netting flows finely and summing them gives the same total as netting once over the whole span, and the window leaves it unchanged. Turnover and market activity are not additive: lengthening $T$ pairs more mints against later burns and shifts offsetting issuance into reshuffling, so $V^E$ generally rises and $V^G$ generally falls. This dependence is intrinsic, not a byproduct of our definition. In a market where the supply of traded claims is endogenous, a mint at time $t_1$ and an offsetting burn at time $t_2 > t_1$ are economically equivalent to an exchange between two traders executed over the interval $[t_1, t_2]$, but Polymarket records them as primary-market activity.

Any measure that separates turnover from issuance must therefore commit to a window and report it. We use monthly $V^E$ for comparisons across periods and daily $V^E$ when the question concerns within-month dynamics, and these two horizons are not directly comparable. However, the horizon choice becomes less critical when one side dominates throughout: October's Trump-market $V^E$ is \$391 million under daily, weekly, or monthly netting alike, because gross burns fall short of gross mints inside every window (\autoref{tab:window_sensitivity}).

Our measures are built on counterparty-matched exchange transactions, in which the matching engine matches traders' orders through exchange, minting, or burning. The settlement contract also supports unilateral operations that a single trader executes against it with no counterparty: direct minting or merging of a complete pair, and, in multi-market events, conversion of a NO portfolio into the economically equivalent YES portfolio plus cash. We exclude all three. A unilateral operation is not counterparty-matched, so it generates no secondary-market turnover and does not affect $V^E$. It is a position-management or arbitrage tool that leaves directional exposure to the event unchanged, so it commits no net directional capital and does not affect $F$. Hence, excluding these three operations omits no trading our measures should count. A trader who merges a pair first bought both legs on the exchange, and those counterparty-matched trades are already counted, so recording the merge too would count the same trading twice. 

The three measures can now be set against the two naive aggregates of \autoref{subsec:trade_mech_hetero}. The \texttt{OrderFilled} aggregate is tied to our measures by an exact accounting identity: (1) a dollar of exchange appears twice in the \texttt{OrderFilled} logs, once in the seller's order and once in the buyer's; (2) a dollar of minting or burning appears once, because the other side of an issuance is the settlement contract, not another trader's order. The aggregate therefore counts exchange twice and issuance once: for any transaction it equals $2\times\text{trade vol} + \text{mint vol} + \text{burn vol}$. Summed over an interval for a single share type, that is twice turnover plus absolute net inflow, $2\,V_{YES}^E(T) + |F_{YES}(T)|$, which is the same quantity as $V_{YES}^E(T) + V_{YES}^G(T)$. Adding the two share types gives the relationship between the naive volume and our measures:
\begin{equation}
\text{naive \texttt{OrderFilled} volume} \;=\; V^E + V^G.
\label{eq:identity}
\end{equation}

\autoref{eq:identity} makes the direction of the bias transparent. The \texttt{OrderFilled} aggregate sums two measures that answer different questions, one asking how many dollars transferred existing risk between traders and the other how large total activity was once capital formation is counted. The aggregate answers neither question. In any window with exchange trading it exceeds both turnover and market activity, so a market read through it looks deeper than it is. This equation also gives the overstatement an exact form. The ratio of naive volume to turnover is $2 + \kappa$, where $\kappa = (V^G - V^E)/V^E$ is the market's issuance-to-turnover ratio: a double-counting floor of two, plus a premium that grows with capital formation and shrinks as trading comes to dominate issuance. We call this ratio the market's \emph{wedge}. \autoref{subsec:crossmarket} uses it to characterize the distortion across the 249-market election-related sample. 

The \texttt{OrdersMatched} aggregate, by contrast, admits no such identity, because the size of the leg it omits moves with the route composition of the interval rather than with any fixed multiple of the three measures. Its distance from our measures is therefore an empirical matter, which the next section reports month by month.

\section{How Large Is the Correction, and How General?}
\label{sec:anatomy}

\noindent This section puts the three measures, turnover $V^E$, net inflow $F$, and market activity $V^G$, to work on the \href{https://polymarket.com/event/presidential-election-winner-2024}{2024 Presidential Election Winner} event in two steps. We first read the three principal candidate markets through both the corrected measures and the naive aggregates, which give different accounts of how the market grew. We then ask whether the gap between them is a quirk of this specific event or a general feature of mint-and-burn settlement.

\subsection{Two accounts of the same cycle}

The corrected measures give one account of the cycle, reported month by month for each candidate market in \autoref{tab:monthly_vol_flow}. The Biden market carries the Democratic side through the first half of 2024 and falls to almost nothing after the withdrawal, while the Harris market rises from \$8 million of turnover in July to \$192 million in October and leads the Trump market on turnover and market activity in August and September. The Trump market then pulls ahead on all three measures in October, the most active month of the sample, recording \$391 million of turnover and \$176 million of net inflow against about half the turnover and 60 percent of the net inflow in the Harris market.

Read through the naive aggregates, the same records give a different account of the same cycle. \autoref{tab:monthly_naive_vol} reports the two naive aggregates introduced in \autoref{subsec:trade_mech_hetero}, each set against turnover. Both aggregates run above turnover. From July to November, in the Harris and Trump markets, the \texttt{OrderFilled} aggregate runs more than twice turnover, and the \texttt{OrdersMatched} aggregate 7 to 39 percent above it. In October the Trump market's \$958 million of naive volume splits into \$391 million of turnover and \$567 million of market activity. The two accounts also disagree about how fast the market grew: from January to October, the Trump market's turnover rose about 459-fold while its naive \texttt{OrderFilled} volume rose only about 266-fold (\autoref{tab:monthly_vol_flow}, \autoref{tab:monthly_naive_vol}), because fresh capital made up much more of the early figure than the later one. The raw figure thus understates how much genuine trading grew, and hides the market's shift from position-building to active trading.

\begin{figure}[hbtp!]
	\begin{subfigure}{\linewidth}
		\centering
		\includegraphics[width=0.8\textwidth]{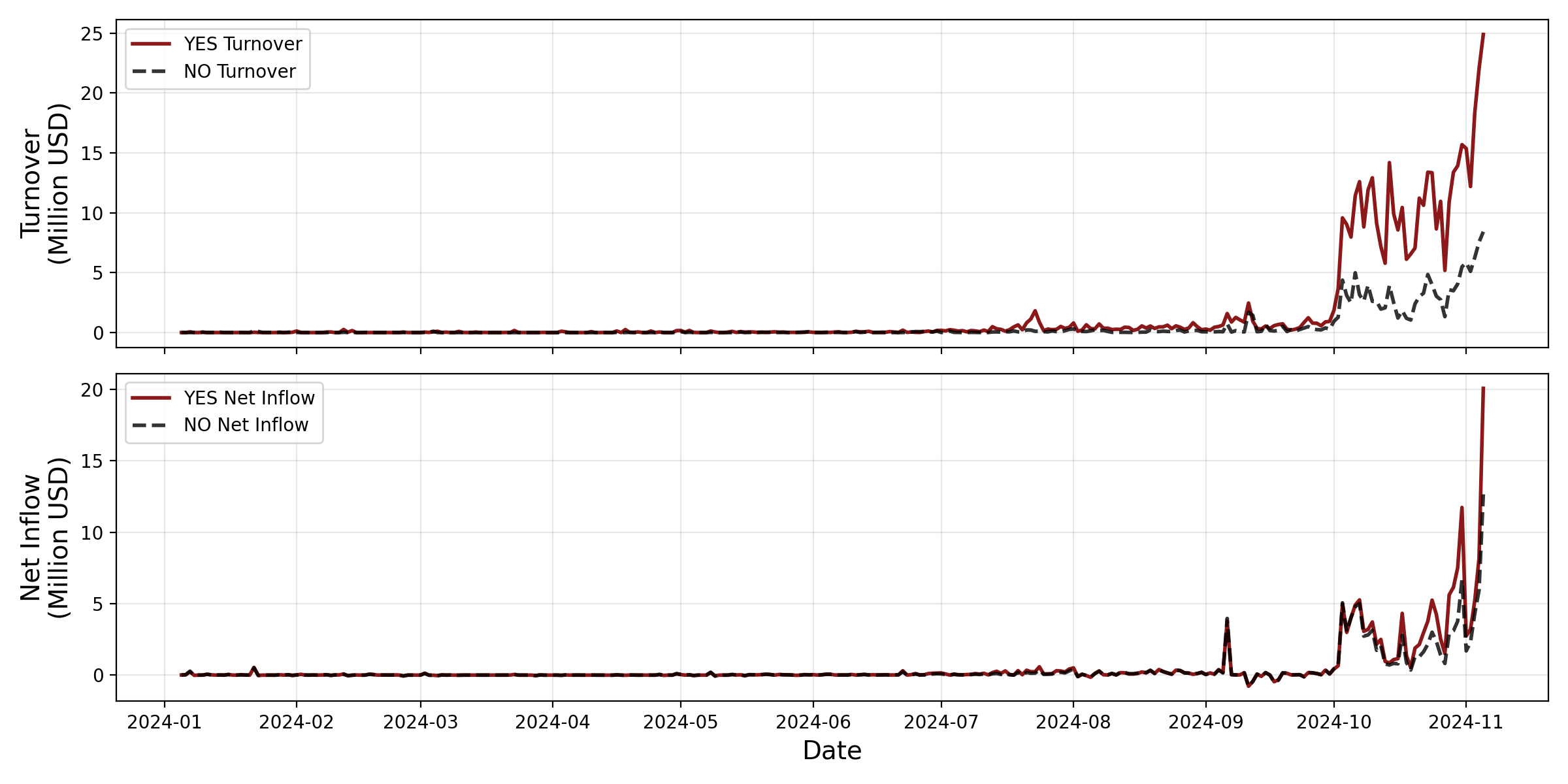}
		\caption{Prediction Market for Donald Trump}
	\end{subfigure}
	\begin{subfigure}{\linewidth}
		\centering
		\includegraphics[width=0.8\textwidth]{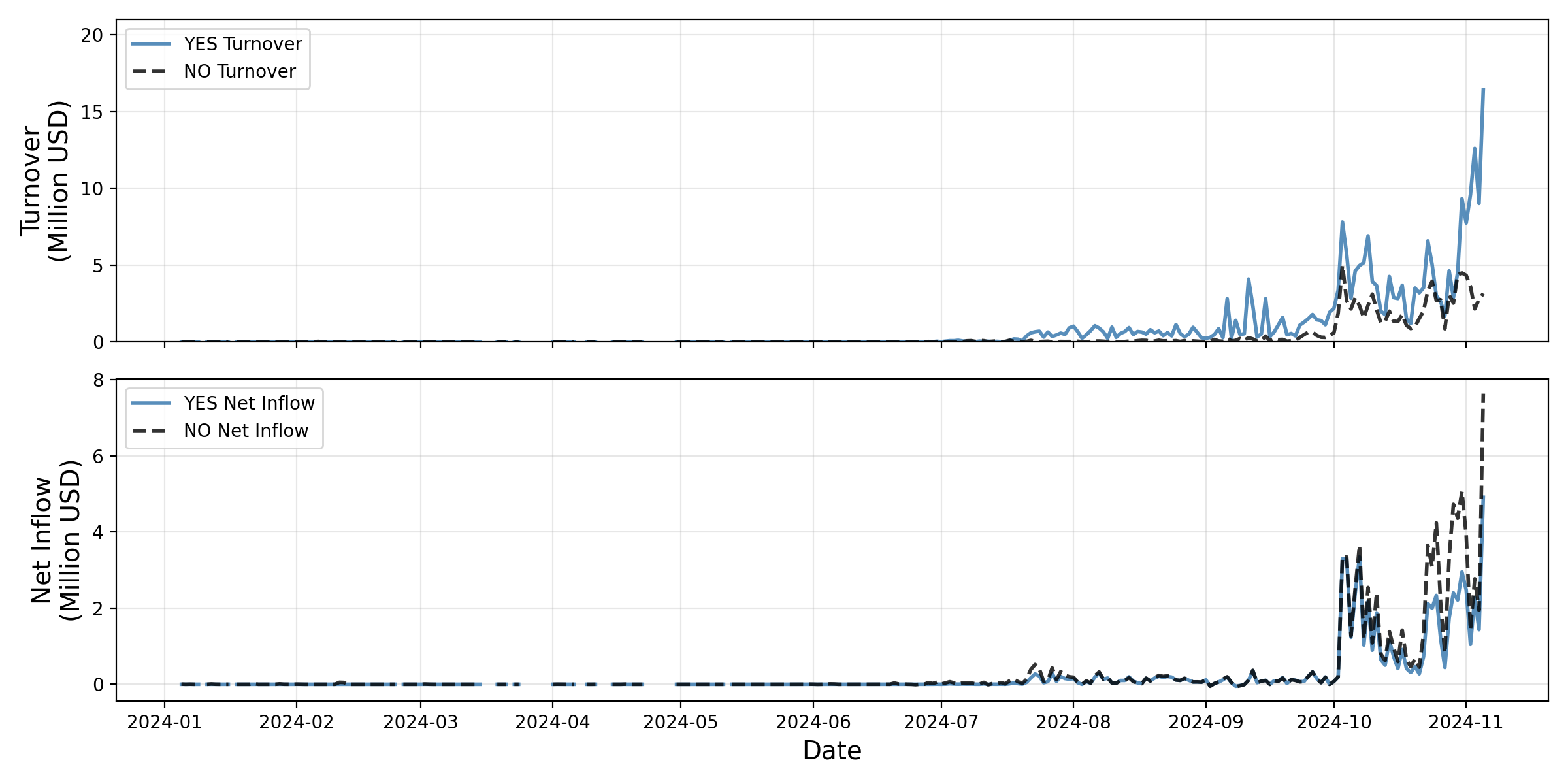}
		\caption{Prediction Market for Kamala Harris}
	\end{subfigure}
	\begin{subfigure}{\linewidth}
		\centering
		\includegraphics[width=0.8\textwidth]{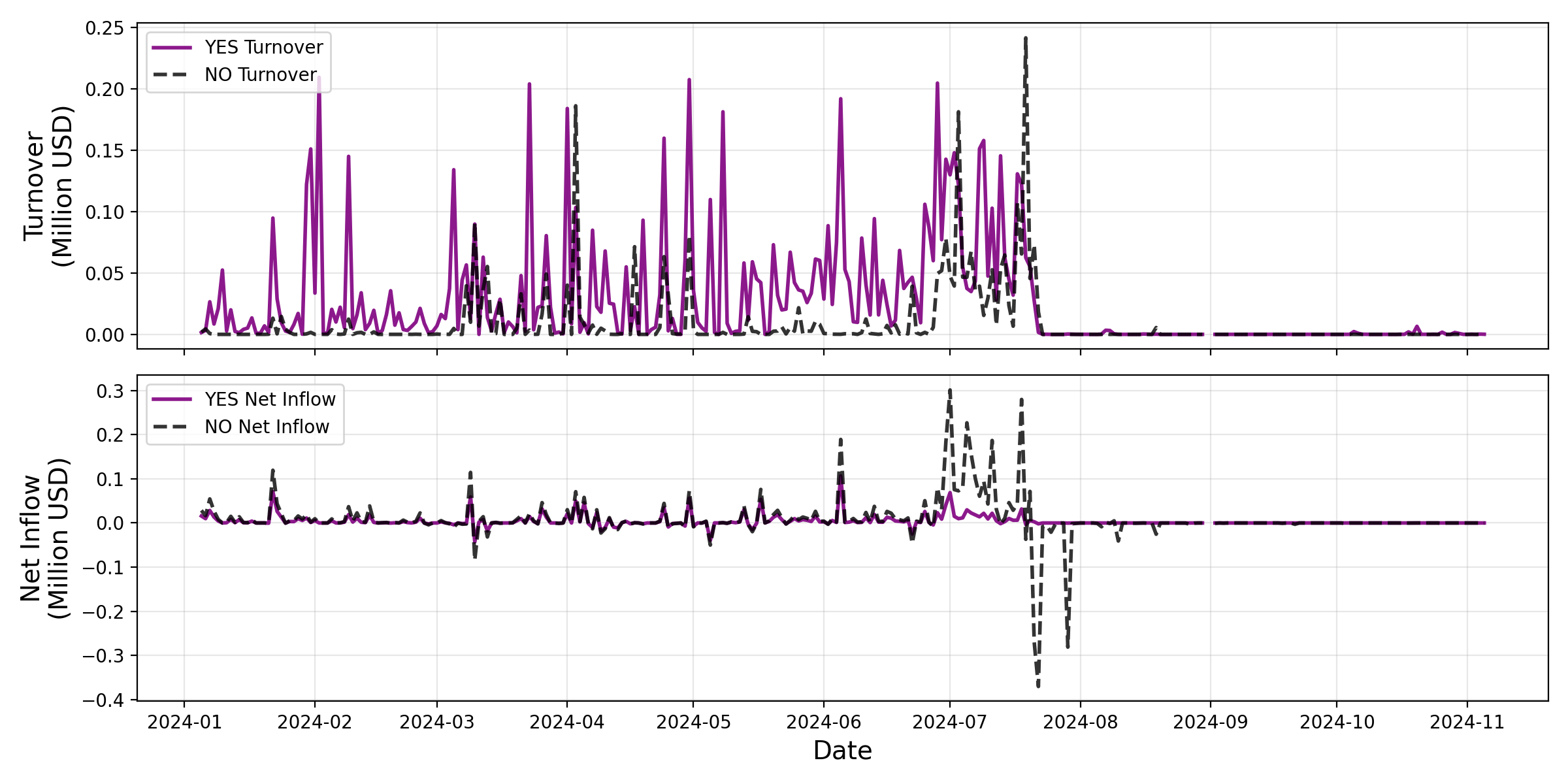}
		\caption{Prediction Market for Joe Biden}
	\end{subfigure}
	\caption{Daily Turnover and Net Inflow of YES and NO Shares}
	\label{fig:daily_volume}
	\vspace{0.5em}
	\begin{minipage}{\linewidth}
		\footnotesize \textit{Note:} This figure plots daily exchange-equivalent turnover ($V^E$, top panel) and net inflow ($F$, bottom panel) for YES and NO shares in each candidate market. Solid lines show YES shares and dashed lines NO shares. The daily series run from January 5 to November 5, 2024, the last full day before the race call.
	\end{minipage}
\end{figure}

The daily view adds a directional signal that monthly totals cannot show. \autoref{fig:daily_volume} plots turnover and net inflow daily for the YES and NO shares of each candidate market. YES shares carry most of the trading in the Trump and Harris markets, and the Trump YES share dominates as the election approaches. In the final days of the sample, net inflow tilts toward Harris NO and Trump YES, two positions that pay in the same state of the world, pointing toward a Trump victory. The gap between naive volume and turnover documented here is a fact about the three candidate markets. \autoref{subsec:crossmarket} asks whether it holds across the hundreds of markets linked to the election.

\subsection{The wedge across 249 markets}
\label{subsec:crossmarket}

Is the wedge, the factor $2 + \kappa$ by which naive volume exceeds turnover (\autoref{eq:identity}), a quirk of the Trump market, or a general implication of mint-and-burn settlement? The answer sets the scope of everything above.

The wedge runs through the entire sample. We apply the decomposition to all 249 markets linked to the 2024 presidential race, 12.9 million orders spanning predictions on candidate, popular vote, electoral margin, state winner, nomination, and exit timing. Because the wedge equals $2 + \kappa$ and the 2 does not vary across markets, it varies only through $\kappa = (V^G - V^E)/V^E$, the issuance-to-turnover ratio. \autoref{tab:crossmarket_dist} reports how the wedge is distributed. Measured over each market's life, the median active market, one with at least \$0.1 million of naive volume, records three times as much naive volume as turnover, and the wedge ranges from just above the double-counting floor of two among large markets to fifty in a small market that traded for only nine days.

Most of these markets share the Trump market's negative-risk settlement, so the wedge could still be an artifact of that format. It is not. ``Will Biden announce resignation by July 31?'' is the only market in a single-market event, settled through the base \texttt{CTFExchange} contract with no negative-risk adapter and no conversion function, the kind of small, short-lived market that platforms list by the hundreds. The decomposition runs unchanged. Over the market's life, naive volume overstates turnover by a factor of 3.26, against 2.41 for the Trump market over its own life (\autoref{tab:monthly_vol_flow}), because minting supplied nearly three quarters of gross buys (\autoref{tab:binary_market}).

In the analogous NASDAQ-NYSE setting, practitioners commonly applied a fixed adjustment factor to reported NASDAQ volume \citep{andersondyl2007}. Whether a single divisor could serve here depends on what predicts $\kappa$. \autoref{tab:crossmarket_reg} shows that, in the market-month panel, $\kappa$ declines by about 12 percent per month of market age in the pooled sample and by about 9 percent with market fixed effects. Conditional on size and lifespan, markets from single-market events show, if anything, lower issuance intensity than markets in multi-market events, so settlement format does not explain the large wedge in the single-market example above. The Trump market shows the same within-market pattern: its wedge starts above four in January 2024 and settles near two and a half as turnover comes to dominate issuance. Hence no single divisor works, and the decomposition computes turnover market by market from each market's own routes.

\section{Market Depth and Price Formation}
\label{results_lambda}

\noindent Polymarket drew public attention because its prediction markets were treated as forecasts, quoted by media and read by campaigns. That forecast role raises the stakes of price manipulation beyond trading profits, since even a temporary distortion shifts the political narrative while it lasts. \citet{rothschild2015trading} document a single Intrade participant sustaining price differences across venues in 2012, and \citet{hansen2004manipulation} report successful manipulation in political stock markets, though \citet{hanson2006manipulation} find that other traders can offset such attempts. In this section, we estimate price impact under naive and corrected order flow and convert it into the capital needed to move the forecast. We then read the same estimate across the negative-risk election markets and ask which layer of activity, secondary-market trading or primary issuance, carries the larger price impact per dollar.

\subsection{Is the market as deep as it looks?}
\label{subsec:lambda_estimation}

We use Kyle's $\lambda$ as a measure of price impact \citep{kyle1985continuous}. Kyle's $\lambda$ links price changes to net order flow: a larger $\lambda$ implies that, for a given amount of signed order flow, prices move more, indicating a shallower market. In such a market, a single trader can shift implied probabilities with comparatively little capital. 

Among the six shares we examine, the Trump YES share is the most actively traded and provides the richest transaction-level record. We therefore estimate Kyle's $\lambda$ for Polymarket's Trump YES share. Our approach follows standard high-frequency microstructure practice, adapted to a binary prediction market.

(1) Net order flow

We do not observe the full limit order book, and instead observe the ledger of matched trades recorded on the Polygon blockchain. The ledger records which side initiated each transaction. Within a matched transaction, exactly one \texttt{OrderFilled} record carries the exchange contract in its \texttt{taker} field. That record belongs to the incoming order, so the wallet that signed it is the initiator, while every other record in the transaction belongs to a resting order. The rule identifies the initiator in all 2,348,771 matched transactions of the Trump market.\footnote{The transaction-level \texttt{OrdersMatched} log also names the initiating wallet in its \texttt{takerOrderMaker} field, and it agrees with the one identified in the above method in every transaction.} We then define signed order flow as
\begin{equation}
	q_n = d_n \cdot v_n,
\end{equation}
where $d_n \in \{+1,-1\}$ is the trade direction and $v_n$ is the dollar size of transaction $n$ in USDC, which we scale to millions of USD before estimation so that $\hat\lambda$ is the price impact per \$1 million. 

The direction $d_n$ comes from the initiator's own order. Because that order can arrive on either leg, we read its direction as exposure to a Trump victory: buying YES or selling NO sets $d_n = +1$, and selling YES or buying NO sets $d_n = -1$. For example, when a mint is initiated by a Trump NO buyer, the new YES shares pass to resting buyers, so the transaction adds YES dollars while the aggressive demand runs against Trump, and the flow is signed negative, because $\lambda$ prices the demand that crossed the spread, not the direction of capital. 

\autoref{sec:volume_measurement} determines what $v_n$ consists of. The initiator's order could be filled against a counterparty's existing shares, against newly minted shares, or by burning shares. Hence, we calculate $v_n$ as:
\begin{equation}
	v_n = \text{YES trade vol}_n + \text{YES mint/burn vol}_n,
\end{equation}
where $\text{YES mint/burn vol}_n$ is whichever of the two is non-zero, since a transaction can combine exchange with minting or with burning, but never with both. The naive alternative instead sums the raw \texttt{OrderFilled} USDC legs of the transaction, which equal $2\,\text{YES trade vol}_n + \text{YES mint/burn vol}_n$, because the exchange component enters through the seller's record and again through the buyer's, while minted or burned dollars enter once\footnote{Of the two naive aggregates, only \texttt{OrderFilled} double-counts the peer-to-peer leg. The \texttt{OrdersMatched} aggregate records the initiator's side once, so it does not overstate order flow. We therefore benchmark the corrected flow against \texttt{OrderFilled} alone.}. Both amounts take the same sign $d_n$, so the corrected and naive flows differ only in $v_n$, and any difference in $\hat\lambda$ comes only from the measure of $v_n$.

The matched trade of \autoref{tab_example_trade_data} makes the difference concrete. In this trade, the \texttt{OrderFilled} record at log index 316 carries the exchange contract \texttt{0xC5d} in its \texttt{taker} field, so its signer, wallet \texttt{0xf0b}, is the initiator, who paid 100.00 USDC for 238.10 Trump YES shares. Two hundred of those shares come from \texttt{0x869}'s holdings, 84.00 USDC of exchange, and the remaining 38.10 are minted, whose YES leg costs 16.00 USDC. The corrected amount is therefore $v_n = 84.00 + 16.00 = 100.00$, the dollars the initiator actually committed, while the naive amount is 184.00 because \texttt{0x869}'s 84.00 USDC appears a second time inside \texttt{0xf0b}'s own record. 

(2) Time aggregation

Prediction-market trading is irregular, so we aggregate trades into fixed intervals before estimating price impact. We use 15-minute intervals. For each interval $t$ we compute the volume-weighted average price (VWAP) $p_t^{\text{VWAP}}$ and net order flow $Q_t$ as
\begin{equation}
	p_t^{\text{VWAP}}=\frac{\sum_{n\in t} w_n p_n}{\sum_{n\in t} w_n}, \qquad
	Q_t=\sum_{n\in t} q_n,
\end{equation}
where $p_n$ is the transaction's average price and $w_n$ its corrected number of Trump YES shares. If no trades occurred in a given interval, we carry forward the last price and set $Q_t=0$ to maintain a continuous time series.

(3) Log-odds price transformation

Because a prediction-market price is bounded between 0 and 1, price changes are naturally heteroskedastic: a five-cent move is more consequential when the price is \$0.10 than when it is \$0.50. To account for this, we convert prices to log odds, which gives an unbounded measure that is more comparable across price levels and reduces nonlinearity near 0 and 1:
\begin{equation}
	\theta_t=\ln \left(\frac{p_t}{1-p_t}\right).
\end{equation}
We then take first differences $\Delta \theta_t = \theta_t - \theta_{t-1}$ as our measure of price movement in each interval, analogous to a return.

(4) Estimation

We estimate Kyle's $\lambda$ by regressing price changes on net order flow,
\begin{equation}
	\Delta \theta_t = \alpha + \lambda \, Q_t + \varepsilon_t,
\end{equation}
with Newey-West standard errors using 96 lags, one trading day of 15-minute intervals. We run this regression for each calendar month, which gives the monthly profile of \autoref{tab:lambda_monthly}, and over 30-day rolling windows stepped forward one day at a time, which gives the series plotted in \autoref{fig:lambda_naive_corr}.

One concern with this regression is that flow responds to past price moves, which would fold feedback trading into $\hat\lambda$. We therefore re-estimate the price response from a monthly vector autoregression in flow and returns (see \autoref{appendix_var}). The dynamic estimates are of the same order as the static ones, so our main results are not driven by feedback trading.

\subsection{The cost of moving the forecast}
\label{subsec:lambda_wedge}

\begin{figure}[hbtp!]
	\centering
	\includegraphics[width=1\textwidth]{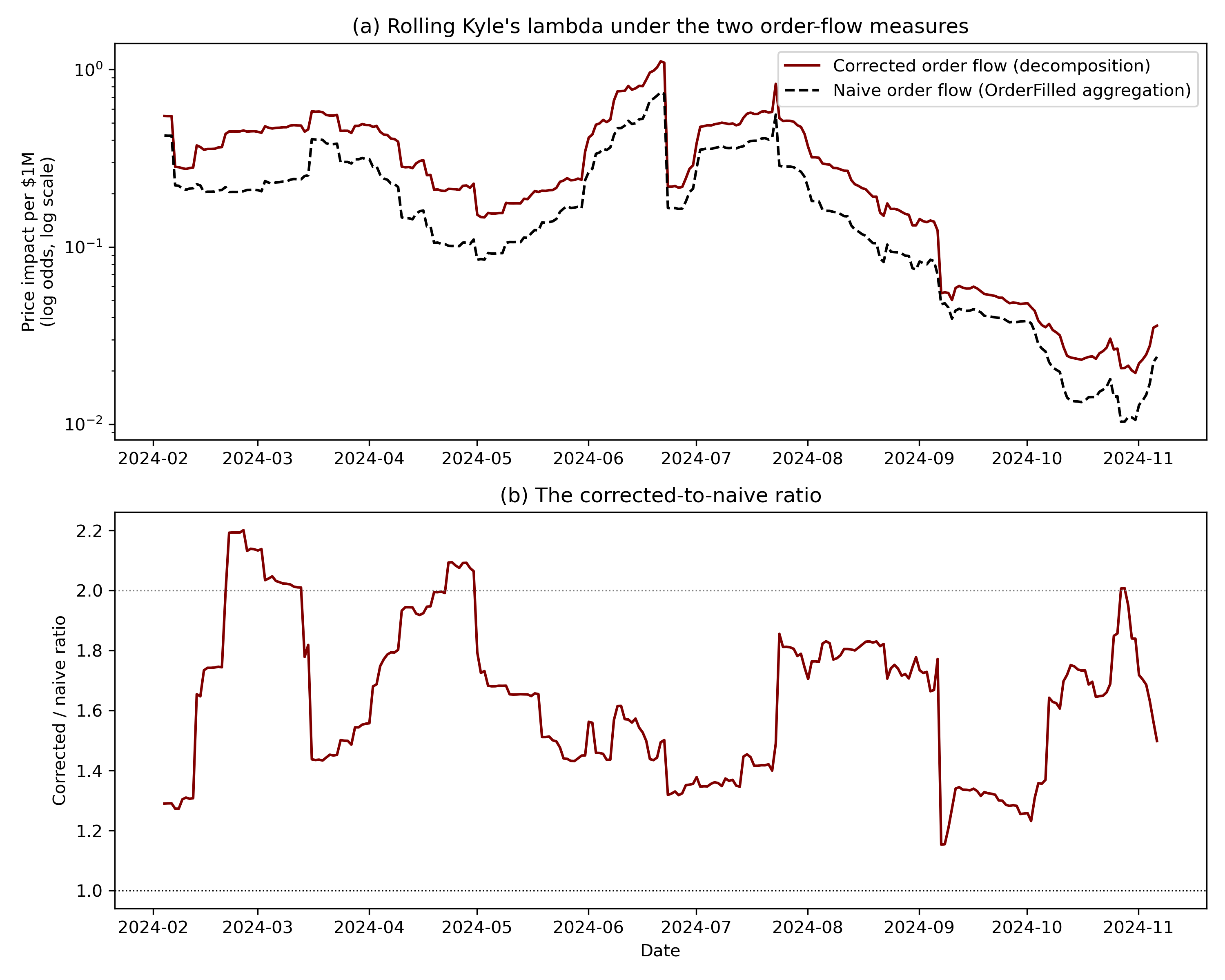}
	\caption{Rolling Kyle's Lambda under Naive and Corrected Order Flow}
	\label{fig:lambda_naive_corr}
	\vspace{0.5em}
	\begin{minipage}{\linewidth}
		\footnotesize \textit{Note:} Panel (a) plots the 30-day rolling estimate of Kyle's $\lambda$ for the Trump YES share on a log scale, from regressions of 15-minute log-odds price changes on 15-minute net signed order flow in million USD over the preceding 30 days with a one-day step. The dashed black line signs the naive \texttt{OrderFilled} flow and the solid maroon line the corrected one-count flow of exchange plus mint plus burn, both using identical prices and the ledger-recorded direction of the initiating order. Panel (b) plots the ratio of the corrected to the naive estimate.
	\end{minipage}
\end{figure}

\autoref{fig:lambda_naive_corr} plots the rolling Kyle's $\lambda$ under naive and corrected order flow. The corrected $\hat\lambda$ lies above the naive one throughout the whole period, and the monthly estimates stand in a ratio of 1.09 to 2.14 in all examined months (\autoref{tab:lambda_monthly}). Naive flow always makes the market look deeper than it was. The ratio is not a constant, ranging from 1.2 to 2.2 across 30-day rolling windows, and it shows no downward drift as the market matures (\autoref{fig:lambda_naive_corr}, Panel (b)): October's monthly ratio of about 1.7 stands above the ten-month average of 1.59. Naive volume therefore overstates the cost of moving the market.

The next question is how costly it actually was to move the market. The corrected column of \autoref{tab:lambda_monthly} shows that price impact falls by more than an order of magnitude as the market deepens over the election cycle. A \$1 million net purchase moves the log-odds price by between 0.15 and 0.59 in the first half of 2024, and by about 0.02 in October. Since $\frac{\mathrm{d}\theta_t}{\mathrm{d}p_t} = \frac{1}{p_t(1-p_t)}$, a first-order expansion converts a log-odds move back into a move in the probability itself,
\begin{equation}
	\Delta \theta_t \approx \frac{1}{p_t(1-p_t)} \Delta p_t \Rightarrow \Delta p_t \approx p_t(1-p_t)\,\Delta \theta_t,
	\label{eq:logodds_prob}
\end{equation}
so that, evaluated at $p_t = 0.5$, a log-odds impact of $\lambda$ per \$1 million is a price move of $0.25\lambda$.\footnote{A purchase moves the price through the $\lambda Q_t$ term alone, so $\alpha$ does not enter the conversion.} The same \$1 million purchase therefore moves the implied probability by roughly 4 to 15 percentage points in the first half of the year, against about half a point in October. By the month in which the price was quoted most widely, the Trump YES share had become substantially less sensitive to order flow.

The difference between the two measures is even sharper in dollars, the capital a trader who wants to move the forecast must actually commit. The cost columns of \autoref{tab:lambda_monthly} report that amount for a move from 0.50 to 0.55. In October it is \$15.6 million under the naive estimate and \$9.1 million under the corrected one. The October market was therefore roughly 40 percent cheaper to move than the naive estimate implies.

\subsection{Depth and price formation across markets}
\label{subsec:crossmarket_depth}

The estimates so far describe the Trump market alone. We turn to the negative-risk election markets, which carry the multi-outcome events and the bulk of the trading, and use the cross-section to examine two things: whether the depth we measured generalizes, and which of the two flows in the corrected $\lambda$ moves the price more per dollar.

\begin{figure}[hbtp!]
	\centering
	\includegraphics[width=0.72\textwidth]{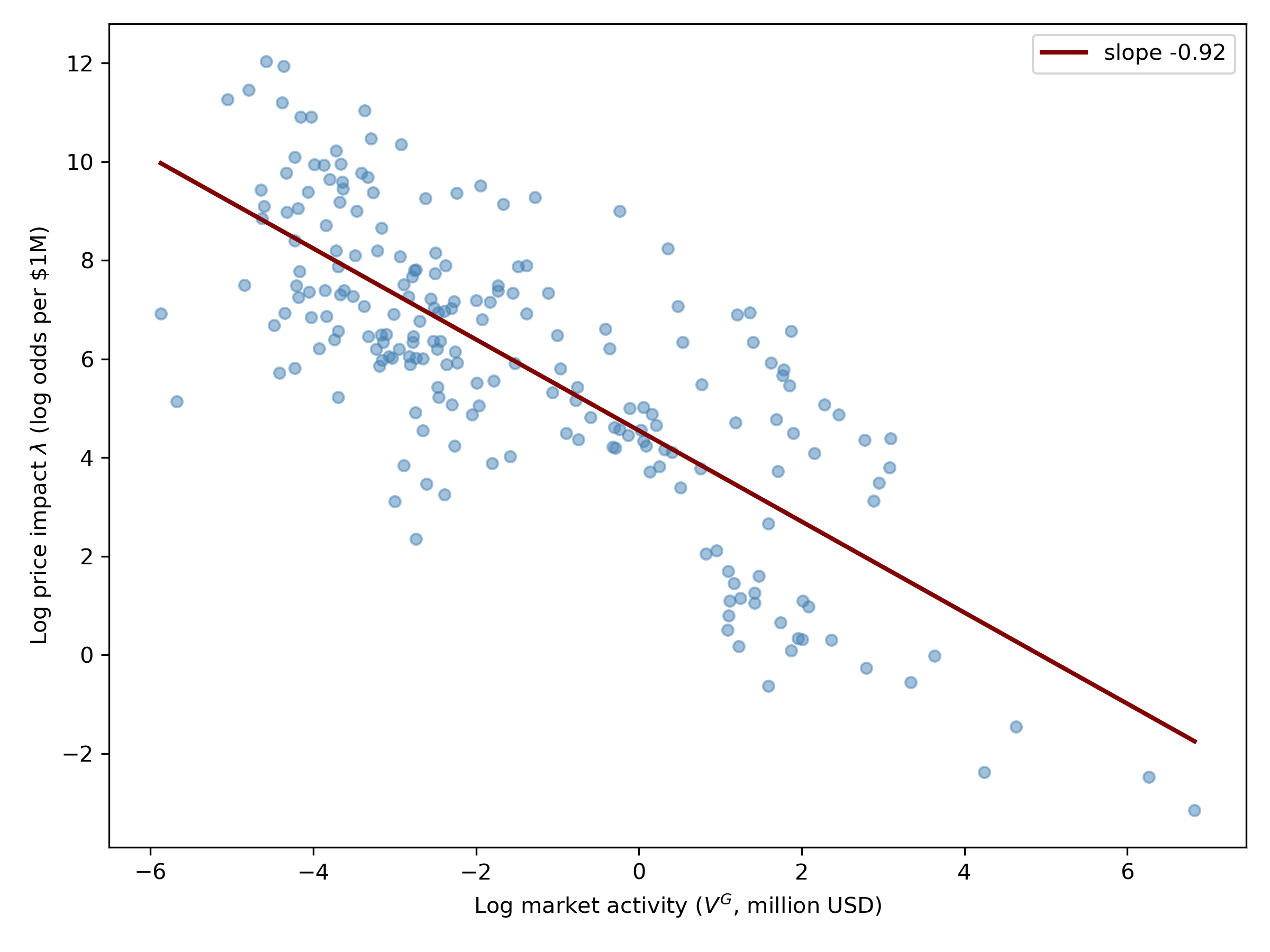}
	\caption{The Depth Gradient Across Election Markets}
	\label{fig:depth_gradient}
	\vspace{0.5em}
	\begin{minipage}{\linewidth}
		\footnotesize \textit{Note:} Each point is one of the 212 negative-risk election markets that have at least 30 fifteen-minute trading bins and a positive estimated $\lambda$, computed from corrected order flow as for the Trump market in \autoref{subsec:lambda_estimation}. Of the 223 negative-risk markets, 10 lack enough bins and one yields a non-positive estimate, and both are excluded from the gradient. The horizontal axis is lifetime market activity $V^G$ and the vertical axis the estimated price impact per \$1 million, both in natural logs. The line is the ordinary least squares fit, with slope $-0.92$ and $R^2 = 0.59$.
	\end{minipage}
\end{figure}

We estimate the corrected $\lambda$ for each of these markets.\footnote{Price impact is estimated from trades placed before the November 6 race call, while market activity is measured over each market's life.} Of these 223 markets, 212 have enough trading bins and a positive estimate and enter the gradient. \autoref{fig:depth_gradient} plots the estimate against market activity $V^G$, both on log scales. Price impact falls with activity at an elasticity of $-0.92$, and activity alone accounts for 59 percent of the variation across these 212 markets.

Meanwhile, on Polymarket, the price forms in two places at once: traders exchange existing shares with one another, the secondary market, and they mint or burn whole contract pairs against the settlement contract, the primary market. Naive volume adds the two together, and even a single corrected $\lambda$ blends them, so neither reveals which flow, trading among existing holders or the issuance of new contracts, has the larger price impact per dollar. Sorting markets into $V^G$ quintiles holds depth roughly fixed, and within each we split the corrected order flow into its secondary and primary components and estimate a price impact for each. \autoref{tab:price_layers} shows that secondary-market flow carries a larger price impact per dollar than primary issuance in every quintile, and the gap is significant in the three more active quintiles, at the one percent level in the third and near the five to ten percent level in the two deepest, where primary issuance is not itself distinguishable from zero.

\section{Who Participated as the Market Deepened?}
\label{sec:traders}

\noindent The cost of moving the forecast is half of the question the market's forecast role raised. The other half is who was trading, and the settlement ledger answers it. In this section we answer three questions. First, who trades in these markets and how their participation develops over the cycle. Second, whether the dollars concentrate among a few accounts as the market deepens. Third, whether capital still enters on both sides of the race even as it concentrates.

\subsection{Participation over the cycle and across markets}
\label{subsec:participation}

Take the arrival question first. \autoref{tab:monthly_participation} counts the addresses trading the three candidate markets month by month. Their number rose from 460 in January 2024 to 139,064 in October, a factor of about 300. Over the same period the three markets' combined activity rose by a factor of about 210. Participation expanded on the extensive margin, and not because a stable group traded more often: 117,234 of October's 139,064 active addresses were trading for the first time.

\begin{landscape}
	\begin{figure}[hbtp!]
		\centering
		\includegraphics[height=0.7\textheight]{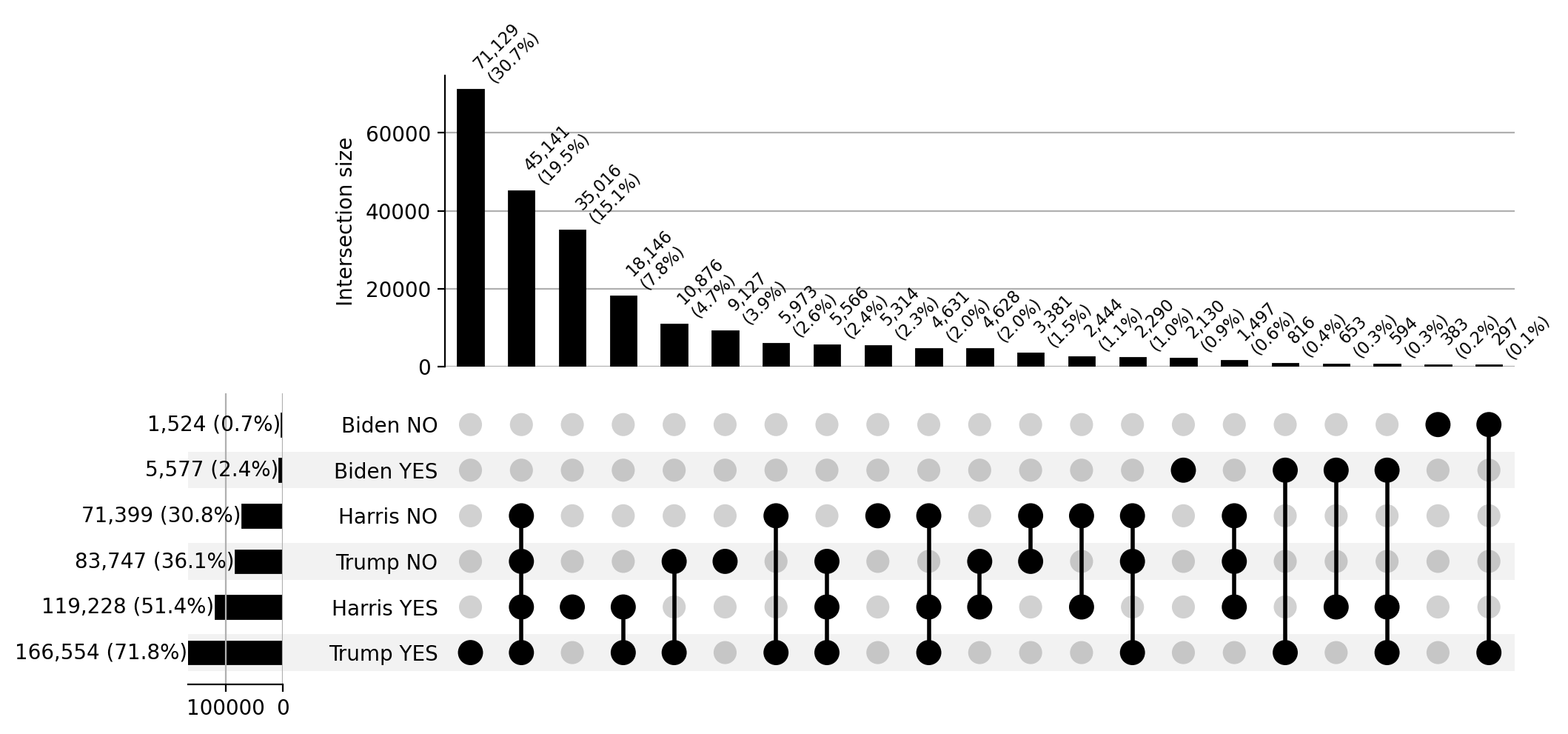}
		\caption{Address Market Participation Distribution}
		\label{fig:upset_trader_participation}
		\vspace{0.5em}
		\begin{minipage}{\linewidth}
			\footnotesize \textit{Note:} This UpSet plot decomposes address participation across all six shares (Trump YES, Trump NO, Harris YES, Harris NO, Biden YES, Biden NO). Each column represents a unique combination of shares in which an address was active, with the bar height giving the number of addresses in the combination and the accompanying label its share of all addresses. The bottom matrix indicates which shares are included in each combination. Combinations covering fewer than 0.1 percent of addresses are omitted.
		\end{minipage}
	\end{figure}
\end{landscape}

Participation is skewed toward the Trump market. \autoref{fig:upset_trader_participation} decomposes address participation across all six shares (Trump YES, Trump NO, Harris YES, Harris NO, Biden YES, and Biden NO). Aggregated back to the candidate level, 39.3 percent of the 231,885 addresses traded only the Trump market, roughly equal to the 41.2 percent that traded in more than one candidate market. This asymmetry mirrors the Trump market's dominant turnover, documented in \autoref{sec:anatomy}.

Across the six shares, three patterns stand out. First, 71.8 percent of all addresses participated in the Trump YES share, and the largest single subgroup, 30.7 percent, consists of addresses that traded exclusively in Trump YES. This concentration is consistent with a large share of participants holding directional exposure favoring a Trump victory, without hedging or trading the opposing side. Second, the next largest subgroup, 19.5 percent, comprises addresses active in all four major shares: Trump YES, Trump NO, Harris YES, and Harris NO. This pattern is consistent with market-making or hedging strategies that require positions across multiple outcomes. Third, addresses that exclusively traded Harris YES constitute the third largest subgroup at 15.1 percent, mirroring the Trump-only pattern on the Democratic side but at roughly half the scale.

\subsection{Concentration at the top}
\label{subsec:whale}

Participation was broad, but the dollars behind it were not evenly spread. October is the natural month to look at, because it is the most active in the sample and the one in which press reports described a single trader placing tens of millions of dollars in pro-Trump bets through four accounts. The ledger lets us go further than the press coverage. We locate those four accounts in the settlement data and match each wallet to its Polymarket display name. We call the trader behind them the whale. As \autoref{tab:whale_concentration} reports, the whale bought \$26.6 million of Trump YES exposure in October. The stake was substantial relative to the month's fresh capital: measured against the \$176.4 million of net inflow that funded the Trump market in October, the whale's purchases equal 15.1 percent.

The whale was not the whole story of the October concentration. Panel B of \autoref{tab:whale_concentration} sets it against every other account, ranking all of them by October net buying with market makers excluded.\footnote{Market makers trade heavily on both sides without carrying exposure, so they would top any gross ranking. Following \citet{tsangyang2026political}, we identify them from the behavioral signature of market making, heavy volume with inventory held near zero \citep{menkveld2013, kirilenko2017}, and remove the 109 accounts that qualify.} October's new capital was concentrated at the top: the largest account bought \$17.0 million net, 9.6 percent of the month's net inflow, the top five 31.1 percent, and the top twenty 50.7 percent. Set against the market's shallowness, this concentration was large: the \$9.1 million needed to move the October forecast five points (\autoref{subsec:lambda_wedge}) was about a third of the whale's net buying and a tenth of the top twenty's. 

\subsection{Both sides of the race}
\label{subsec:disagreement}

A small group of large buyers on one side raises a question about what the Trump price represented: the market's read on the race, or those buyers' conviction. The answer depends on what the rest of the market did. If fresh capital flowed only toward Trump, the price rested on one-sided demand. If other traders were committing fresh capital to the Democratic side at the same time, the race stayed actively contested. We ask which pattern held, first in the aggregate daily flows and then across individual accounts.

\begin{figure}[hbtp!]
	\centering
	\includegraphics[width=1\textwidth]{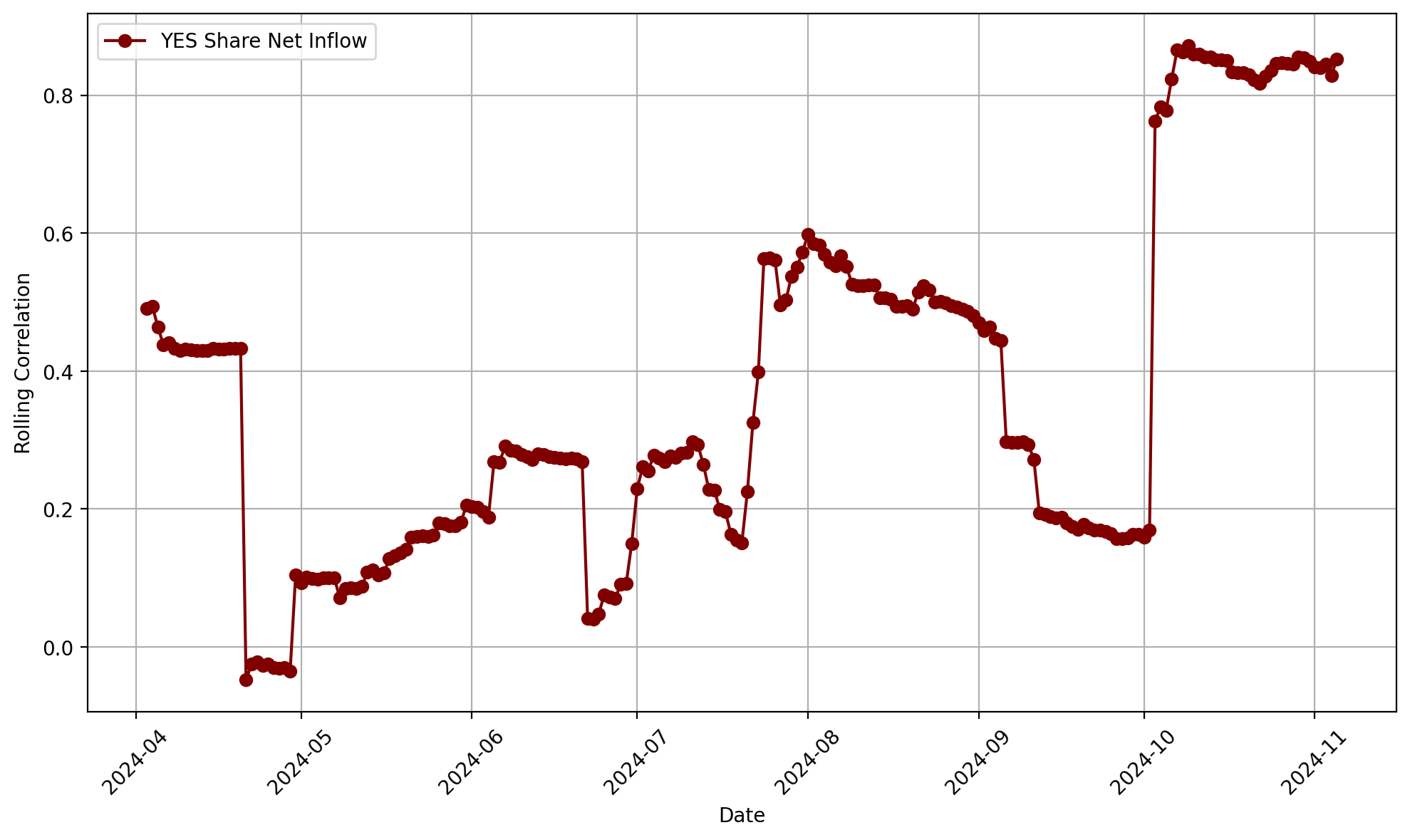}
	\caption{Rolling Correlation Between Trump and Democratic-Side Prediction Markets}
	\label{fig:rolling_correlation_trump_and_dem_market}
	\vspace{0.5em}
	\begin{minipage}{\linewidth}
		\footnotesize \textit{Note:} This figure shows the 90-day rolling correlation of daily YES-share net inflow between the Trump market and the combined Democratic-side markets, with a step size of one day. Higher correlation indicates stronger co-movement in the two daily net-inflow series.
	\end{minipage}
\end{figure}

Start with the aggregate flows. Had the Democratic side gone quiet while capital entered the Trump market, its fresh capital would have flattened and stopped tracking the Trump net inflow. We pool the Biden and Harris markets as the Democratic side, sum their daily YES-share net inflows, and plot the 90-day rolling correlation with Trump YES net inflow (\autoref{fig:rolling_correlation_trump_and_dem_market}). A higher correlation means the two net-inflow series co-moved, so the Democratic side kept drawing fresh capital when the Trump market did.

The correlation moves three times over the cycle. It rises around Biden's withdrawal and Harris's entry, when the nominee switch created uncertainty about how the Democratic ticket would affect the race and the two daily net-inflow series moved together more closely. It falls through September as the two series decoupled, a period that includes the Trump-Harris debate on September 10, 2024, when Polymarket odds swung from a clear Trump lead to a near tie within hours. It then rises through early October and stays above 0.8 from October 6 to the race call, as the concentrated pro-Trump buying of \autoref{subsec:whale} began, so the Democratic side drew fresh capital most closely in step with that buying exactly as it concentrated.

The correlation is only an aggregate, and it cannot reveal its own composition: each side could be populated by many traders or dominated by a few large accounts. Whether both sides were broadly populated is a separate question, one only the cross-section of individual accounts can answer.

To distinguish these cases, we move from aggregate flows to the order-level ledger. We classify each order as pro-Trump or Democratic-side by share type and trade direction and aggregate each address's signed net exposure for the day, counted positive on the pro-Trump side and negative on the Democratic side. We then ask three questions of the daily cross-section: was exposure concentrated in a few large accounts, did active traders split across both sides of the race, and were the dollars split across them too? Exposure dispersion $D_t$, the standard deviation of these net exposures across all active addresses standardized by their mean absolute size, answers the first, headcount polarization $P^N_t = 2\min(N^+_t, N^-_t)/(N^+_t + N^-_t)$ the second, and volume-weighted polarization $P^V_t = 2\min(D^+_t, D^-_t)/(D^+_t + D^-_t)$ the third, with $D^+_t$ and $D^-_t$ the gross dollar volume traded by the pro-Trump and Democratic-side addresses.\footnote{An address is active on day $t$ if its absolute net exposure exceeds one USDC. $N^+_t$ and $N^-_t$ count active addresses with pro-Trump and Democratic-side net exposure.} 

\begin{figure}[hbtp!]
	\centering
	\includegraphics[width=0.9\linewidth]{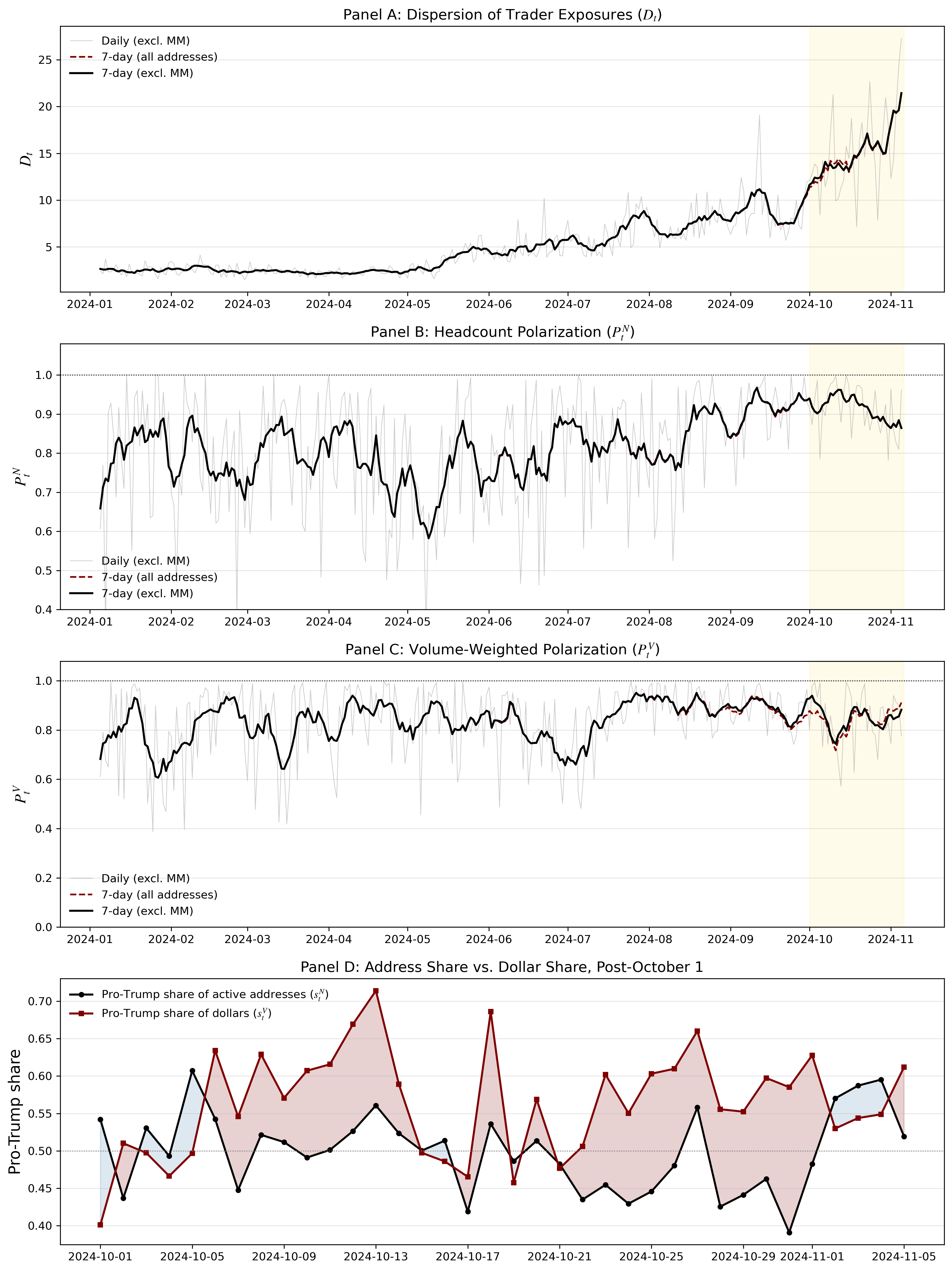}
	\caption{Address-Level Flow Composition Measures}
	\label{fig:disagreement_trader_level}
	\vspace{0.5em}
	\begin{minipage}{\linewidth}
		\footnotesize \textit{Note:} Panels A to C plot the address-level flow composition measures defined in \autoref{subsec:disagreement}, exposure dispersion $D_t$, headcount polarization $P^N_t$, and volume-weighted polarization $P^V_t$. Panel D plots the pro-Trump share of active addresses and of dollars traded after October~1, 2024. In Panels A to C, the thin gray line is the daily series and the thick black line its 7-day centered moving average, both excluding the 109 behavioral market makers. The dashed maroon line is the all-address 7-day average. Panel D likewise excludes the flagged accounts. The gold-shaded region marks the period after October 1.
	\end{minipage}
\end{figure}

\autoref{fig:disagreement_trader_level} plots the four panels.\footnote{As in \autoref{subsec:whale}, these series exclude the behavioral market makers.} Exposure dispersion $D_t$ (Panel A) rises sharply after mid-September and continues increasing through October, as the cross-sectional distribution of daily net exposures becomes more unequal with the entry of the large directional accounts of \autoref{subsec:whale}. Rising concentration turns a market one-sided only if the other side retreats. Panels B and C show it did not. Headcount polarization $P^N_t$ remains high around the start of October and declines only in the final weeks, and volume-weighted polarization $P^V_t$ is high across most of the window. Both sides kept comparable numbers of traders and, for most of the month, comparable dollars, so the concentrated buying did not leave the market one-sided.

$P^V_t$ does move once. It falls in early October, when dollar volume became temporarily concentrated on one side as the large-account episode began, and rises again later in the month as additional Democratic-side volume entered against the large pro-Trump positions. Panel D reconciles that fall with the steadier headcount series: in late October, addresses with Democratic-side exposure often outnumber those with pro-Trump exposure, but the pro-Trump side accounts for a larger share of dollar volume than of headcount, an increasingly asymmetric population with larger average trade size on the pro-Trump side.

October's Trump price therefore emerged from a contested market. The buying was concentrated, as \autoref{subsec:whale} showed and the rising dispersion $D_t$ confirms, but concentration among a few accounts is not the same as a one-sided market. Across the two sides of the race, both stayed populated and funded as the large buyers accumulated, much of the activity coming from directional addresses each committed to one side of one candidate. Fresh capital kept entering on the Democratic side as the pro-Trump buying concentrated, so the forecast reflected a two-sided contest under heterogeneous beliefs \citep{harris1993differences, hong2007disagreement} rather than one-sided demand.

\section{Conclusion}
\label{conclusion}

\noindent In a market where claims are minted and burned on demand, trading volume is constructed rather than directly observed. Naive aggregation of the settlement records combines secondary-market turnover with primary issuance and redemption, so it answers neither the depth question nor the capital-formation question cleanly. This paper develops a transaction-level decomposition that separates turnover from net inflow and market activity, and shows that replacing the naive aggregate with these measures changes what the records say about the market, not merely the magnitudes: how much its trading grew, and how costly its price was to move.

The same decomposition applies to any venue that settles by minting and burning collateralized outcome shares and whose records reveal each transaction's route. Across the 249-market election-related sample, the overstatement runs from its floor of two to a factor of fifty, largest in thin and young markets and converging toward the floor as markets mature. The correction is therefore a general implication of mint-and-burn settlement, not a quirk of the market where we found it.

The Trump market shows what the correction changes. Naive volume made it look far deeper, and so harder to move, than it was: moving the October forecast by five percentage points cost about \$9.1 million of corrected order flow, not the \$15.6 million the raw aggregate implied. In so shallow a market, the concentration of October's buying could have made the forecast the work of a few accounts: the reported whale bought about 15 percent of net inflow, and the twenty largest buyers about half. Yet capital kept entering on both sides of the race as they accumulated, so the forecast was not the work of the concentrated few alone.

The welfare implications of prediction market activity remain open. Whether these markets improve forecasting, distort information, or create new channels for political influence is a separate question, but the measurement framework developed here provides a basis for studying it.

\clearpage

\onehalfspacing

\bibliographystyle{chicago}

\bibliography{polymarket_bib}

@article{andersondyl2007,
  title={Trading volume: {NASDAQ} and the {NYSE}},
  author={Anderson, Anne M and Dyl, Edward A},
  journal={Financial Analysts Journal},
  volume={63},
  number={3},
  pages={79--86},
  year={2007}
}

@article{huangstoll1996,
  title={Dealer versus auction markets: A paired comparison of execution costs on {NASDAQ} and the {NYSE}},
  author={Huang, Roger D and Stoll, Hans R},
  journal={Journal of Financial Economics},
  volume={41},
  number={3},
  pages={313--357},
  year={1996}
}

@article{kyle1985continuous,
  title={Continuous auctions and insider trading},
  author={Kyle, Albert S},
  journal={Econometrica},
  volume={53},
  number={6},
  pages={1315--1335},
  year={1985},
  doi={10.2307/1913210}
}

@article{wolfers2004prediction,
  title={Prediction markets},
  author={Wolfers, Justin and Zitzewitz, Eric},
  journal={Journal of Economic Perspectives},
  volume={18},
  number={2},
  pages={107--126},
  year={2004}
}

@techreport{wolfers2006interpreting,
  title={Interpreting prediction market prices as probabilities},
  author={Wolfers, Justin and Zitzewitz, Eric},
  year={2006},
  institution={National Bureau of Economic Research},
  type={Working Paper},
  number={12200}
}

@article{manski2006interpreting,
  title={Interpreting the predictions of prediction markets},
  author={Manski, Charles F},
  journal={Economics Letters},
  volume={91},
  pages={425--429},
  year={2006}
}

@article{rothschild2015trading,
  title={Trading strategies and market microstructure: Evidence from a prediction market},
  author={Rothschild, David M and Sethi, Rajiv},
  journal={Journal of Prediction Markets},
  volume={10},
  number={1},
  pages={1--29},
  year={2016}
}

@inproceedings{chen2024political,
  title={Political leanings in {Web3} betting: Decoding the interplay of political and profitable motives},
  author={Chen, Hongzhou and Duan, Xiaolin and El Saddik, Abdulmotaleb and Cai, Wei},
  booktitle={Proceedings of the 17th {ACM} Web Science Conference 2025},
  pages={96--105},
  year={2025},
  publisher={Association for Computing Machinery},
  doi={10.1145/3717867.3717884}
}

@unpublished{ng2026price,
  title={Price discovery and trading in modern prediction markets},
  author={Ng, Hunter and Peng, Lin and Tao, Yubo and Zhou, Dexin},
  year={2026},
  note={SSRN Working Paper No.~5331995}
}

@unpublished{rahman2026sok,
  title={{SoK}: Market microstructure for decentralized prediction markets ({DePMs})},
  author={Rahman, Nahid and Al-Chami, Joseph and Clark, Jeremy},
  year={2026},
  note={Workshop on Trusted Smart Contracts (WTSC), Financial Cryptography 2026; arXiv:2510.15612},
  doi={10.48550/arXiv.2510.15612}
}

@article{arrow2008promise,
  title={The promise of prediction markets},
  author={Arrow, Kenneth J and Forsythe, Robert and Gorham, Michael and Hahn, Robert and Hanson, Robin and Ledyard, John O and Levmore, Saul and Litan, Robert and Milgrom, Paul and Nelson, Forrest D and others},
  journal={Science},
  volume={320},
  number={5878},
  pages={877--878},
  year={2008}
}

@article{harris1993differences,
  title={Differences of opinion make a horse race},
  author={Harris, Milton and Raviv, Artur},
  journal={The Review of Financial Studies},
  volume={6},
  number={3},
  pages={473--506},
  year={1993}
}

@article{hong2007disagreement,
  title={Disagreement and the stock market},
  author={Hong, Harrison and Stein, Jeremy C},
  journal={Journal of Economic Perspectives},
  volume={21},
  number={2},
  pages={109--128},
  year={2007}
}

@article{allen1992stock,
  title={Stock-price manipulation},
  author={Allen, Franklin and Gale, Douglas},
  journal={The Review of Financial Studies},
  volume={5},
  number={3},
  pages={503--529},
  year={1992}
}

@article{hanson2006manipulation,
  title={Information aggregation and manipulation in an experimental market},
  author={Hanson, Robin and Oprea, Ryan and Porter, David},
  journal={Journal of Economic Behavior \& Organization},
  volume={60},
  number={4},
  pages={449--459},
  year={2006}
}

@article{hansen2004manipulation,
  title={Manipulation in political stock markets---preconditions and evidence},
  author={Hansen, Jan and Schmidt, Carsten and Strobel, Martin},
  journal={Applied Economics Letters},
  volume={11},
  number={7},
  pages={459--463},
  year={2004}
}

@techreport{sirolly2025network,
  title={Network-Based Detection of Wash Trading},
  author={Sirolly, Allen and Ma, Hongyao and Kanoria, Yash and Sethi, Rajiv},
  institution={Columbia Business School},
  type={Working Paper},
  number={5714122},
  year={2025},
  doi={10.2139/ssrn.5714122}
}

@misc{slivkoff2025paradigm,
  title={Polymarket volume is being double-counted},
  author={Slivkoff, Storm},
  year={2025},
  howpublished={Paradigm Research},
  note={December 2025}
}

@article{berg2008accuracy,
  title={Prediction market accuracy in the long run},
  author={Berg, Joyce E and Nelson, Forrest D and Rietz, Thomas A},
  journal={International Journal of Forecasting},
  volume={24},
  number={2},
  pages={285--300},
  year={2008}
}

@article{menkveld2013,
  author  = {Menkveld, Albert J.},
  title   = {High Frequency Trading and the New Market Makers},
  journal = {Journal of Financial Markets},
  year    = {2013},
  volume  = {16},
  number  = {4},
  pages   = {712--740}
}

@article{kirilenko2017,
  author  = {Kirilenko, Andrei A. and Kyle, Albert S. and Samadi, Mehrdad and Tuzun, Tugkan},
  title   = {The Flash Crash: High-Frequency Trading in an Electronic Market},
  journal = {Journal of Finance},
  year    = {2017},
  volume  = {72},
  number  = {3},
  pages   = {967--998}
}

@unpublished{tsangyang2026political,
  author = {Tsang, Kwok Ping and Yang, Zichao},
  title  = {Political Shocks and Price Discovery in Prediction Markets: Evidence from the 2024 {U.S.} Presidential Election},
  year   = {2026},
  note   = {arXiv preprint arXiv:2603.03152}
}

\clearpage



\clearpage

\clearpage


\begin{table}[hbtp!]
	\centering
	\caption{Examples of Filled Orders}
	\label{tab_example_orders_types}
	\normalsize
	\setlength{\tabcolsep}{2.5pt}
	\renewcommand{\arraystretch}{1.1}
	\resizebox{\textwidth}{!}{%
	\begin{tabular}{lccccccccc}
		\hline
		\hline
		Type & block & txIndex & logIndex & maker & taker & makerAssetId & takerAssetId & makerAmountFilled & takerAmountFilled \\
		\midrule
		\multirow{3}{*}{Simple Trade} & 51953200 & 180 & 821 & 0x869... & 0x9d8... & Trump NO (483310...) & 0 & 200 & 118 \\
		& 51953200 & 180 & 824 & 0xd42... & 0x9d8... & Trump NO (483310...) & 0 & 10 & 5.9 \\
		& 51953200 & 180 & 826 & 0x9d8... & 0xC5d... & 0 & Trump NO (483310...) & 123.9 & 210 \\
		\midrule
		\multirow{2}{*}{Share Minting} & 54432034 & 44 & 137 & 0x351... & 0xE0D... & 0 & Biden NO (347316...) & 3960 & 6000 \\
		& 54432034 & 44 & 139 & 0xE0D... & 0xC5d... & 0 & Biden YES (880278...) & 2040 & 6000 \\
		\midrule
		\multirow{2}{*}{Share Burning} & 51958552 & 74 & 293 & 0x64C... & 0xff6... & Trump NO (483310...) & 0 & 206.19 & 123.714 \\
		& 51958552 & 74 & 295 & 0xff6... & 0xC5d... & Trump YES (217426...) & 0 & 206.19 & 82.476 \\
		\midrule
		\multirow{3}{*}{Mixed Trade} & 51951654 & 70 & 300 & 0x869... & 0xf0b... & Trump YES (217426...) & 0 & 200 & 84 \\
		& 51951654 & 70 & 314 & 0xd42... & 0xf0b... & 0 & Trump NO (483310...) & 22.095238 & 38.095237 \\
		& 51951654 & 70 & 316 & 0xf0b... & 0xC5d... & 0 & Trump YES (217426...) & 99.999999 & 238.095237 \\
		\hline
		\hline
	\end{tabular}%
	}
	\vspace{0.5em}
	\begin{minipage}{\linewidth}
		\footnotesize \textit{Note:} This table presents representative \texttt{OrderFilled} logs from the 2024 Presidential Election Winner event on Polymarket, illustrating the four trade types: simple trade, share minting, share burning, and mixed trade. Each amount is in its asset's own units, shares for an outcome-share leg and USDC for the asset-\texttt{0} leg, rescaled from the raw on-chain values recorded in $10^{-6}$ units. Wallet addresses and asset IDs are truncated for readability. 
	\end{minipage}
\end{table}

\begin{table}[hbtp!]
	\centering
	\caption{Selected Event-Log Fields for a Matched Trade}
	\label{tab_example_trade_data}
	\normalsize
	\setlength{\tabcolsep}{2.5pt}
	\renewcommand{\arraystretch}{1.1}
	\resizebox{\textwidth}{!}{%
	\begin{tabular}{lccccccccc}
		\hline
		\hline
		block & txIndex & logIndex & maker & taker & makerAssetId & takerAssetId & makerAmountFilled & takerAmountFilled & event \\
		\midrule
		51951654 & 70 & 300 & 0x869... & 0xf0b... & Trump YES (217426...) & 0 & 200 & 84 & OrderFilled \\
		51951654 & 70 & 314 & 0xd42... & 0xf0b... & 0 & Trump NO (483310...) & 22.095238 & 38.095237 & OrderFilled \\
		51951654 & 70 & 316 & 0xf0b... & 0xC5d... & 0 & Trump YES (217426...) & 99.999999 & 238.095237 & OrderFilled \\
		\midrule
		51951654 & 70 & 317 &  &  & 0 & Trump YES (217426...) & 99.999999 & 238.095237 & OrdersMatched \\
		\hline
		\hline
	\end{tabular}%
	}
	\vspace{0.5em}
	\begin{minipage}{\linewidth}
		\footnotesize \textit{Note:} This table presents selected fields from the logs of the Mixed Trade in \autoref{tab_example_orders_types}. The log contains both \texttt{OrderFilled} and \texttt{OrdersMatched} logs. Wallet addresses and asset IDs are truncated for readability. Each asset ID appears as the outcome share it denotes followed by the first six digits of its token ID, and an asset ID of \texttt{0} denotes USDC.
	\end{minipage}
\end{table}


\begin{table}[htbp!]
	\centering
	\caption{Monthly Turnover, Net Inflow, and Market Activity by Candidate (Million USD)}
	\label{tab:monthly_vol_flow}
	\setlength{\tabcolsep}{4pt}
	\renewcommand{\arraystretch}{1.1}
	\resizebox{\textwidth}{!}{%
	\begin{tabular}{cccccccccc}
	\hline
	\hline
	 & \multicolumn{3}{c}{Turnover ($V^E$)} & \multicolumn{3}{c}{Net Inflow ($F$)} & \multicolumn{3}{c}{Market Activity ($V^G$)} \\
	\cmidrule(lr){2-4} \cmidrule(lr){5-7} \cmidrule(lr){8-10}
	Month & Biden & Harris & Trump & Biden & Harris & Trump & Biden & Harris & Trump \\
	\midrule
	1  & 0.654  & 0.010  & 0.852   & 0.652   & 0.053   & 1.904   & 1.306   & 0.063   & 2.756   \\
	2  & 0.703  & 0.053  & 1.199   & 0.246   & 0.109   & 0.427   & 0.949   & 0.162   & 1.626   \\
	3  & 1.557  & 0.016  & 1.029   & 0.200   & 0.010   & 0.235   & 1.757   & 0.026   & 1.264   \\
	4  & 1.924  & 0.010  & 1.368   & 0.394   & 0.011   & 0.263   & 2.318   & 0.021   & 1.631   \\
	5  & 1.331  & 0.059  & 2.171   & 0.314   & 0.006   & 1.036   & 1.646   & 0.064   & 3.207   \\
	6  & 2.120  & 0.174  & 2.479   & 0.946   & 0.197   & 2.316   & 3.066   & 0.371   & 4.795   \\
	7  & 4.063  & 8.276  & 13.747  & 1.215   & 5.495   & 7.884   & 5.278   & 13.771  & 21.631  \\
	8  & 0.021  & 20.930 & 16.877  & $-$0.077 & 7.214   & 8.920   & 0.098   & 28.145  & 25.797  \\
	9  & 0.001  & 40.833 & 35.859  & $-$0.006 & 5.528   & 8.249   & 0.007   & 46.361  & 44.108  \\
	10 & 0.017  & 191.928 & 391.030 & $-$0.002 & 106.229 & 176.417 & 0.019   & 298.157 & 567.448 \\
	11 & 0.001  & 92.391 & 175.538  & $-$0.000 & 44.087  & 57.807  & 0.001   & 136.478  & 233.345 \\
	\hline
	\hline
	\end{tabular}%
	}
	\vspace{0.5em}
	\begin{minipage}{\linewidth}
		\footnotesize \textit{Note:} This table reports monthly exchange-equivalent turnover ($V^E$), net inflow ($F$), and gross market activity ($V^G$) for the Biden, Harris, and Trump prediction markets. Each measure sums the market's YES and NO shares. The sample ends on November~6, 2024, so month~11 contains fewer than six trading days. All values are in millions of USD.
	\end{minipage}
\end{table}


\begin{table}[htbp!]
	\centering
	\caption{Monthly Naive Trading Volume Aggregates (Million USD)}
	\label{tab:monthly_naive_vol}
		\setlength{\tabcolsep}{4pt}
		\renewcommand{\arraystretch}{1.05}
		\begin{tabular}{ccccccc}
	\hline
	\hline
	 & \multicolumn{3}{c}{Use \texttt{OrdersMatched} logs} & \multicolumn{3}{c}{Use \texttt{OrderFilled} logs} \\
	\cmidrule(lr){2-4} \cmidrule(lr){5-7}
	Month & Biden & Harris & Trump & Biden & Harris & Trump \\
	\midrule
		1  & 0.989 & 0.061 & 1.806 & 1.960 & 0.073 & 3.608 \\
		   & (1.51) & (6.29) & (2.12) & (3.00) & (7.47) & (4.23) \\
		2  & 0.862 & 0.091 & 1.444 & 1.652 & 0.216 & 2.824 \\
		   & (1.23) & (1.70) & (1.20) & (2.35) & (4.05) & (2.36) \\
		3  & 1.628 & 0.023 & 1.158 & 3.314 & 0.043 & 2.294 \\
		   & (1.05) & (1.43) & (1.12) & (2.13) & (2.61) & (2.23) \\
		4  & 2.059 & 0.018 & 1.482 & 4.242 & 0.031 & 2.999 \\
		   & (1.07) & (1.83) & (1.08) & (2.20) & (3.14) & (2.19) \\
		5  & 1.491 & 0.102 & 2.713 & 2.977 & 0.123 & 5.378 \\
		   & (1.12) & (1.74) & (1.25) & (2.24) & (2.10) & (2.48) \\
		6  & 2.670 & 0.326 & 3.762 & 5.186 & 0.545 & 7.274 \\
		   & (1.26) & (1.87) & (1.52) & (2.45) & (3.13) & (2.93) \\
		7  & 4.906 & 11.525 & 18.240 & 9.341 & 22.046 & 35.378 \\
		   & (1.21) & (1.39) & (1.33) & (2.30) & (2.66) & (2.57) \\
		8  & 0.098 & 24.524 & 21.430 & 0.119 & 49.075 & 42.675 \\
		   & (4.65) & (1.17) & (1.27) & (5.65) & (2.34) & (2.53) \\
		9  & 0.007 & 43.586 & 39.938 & 0.007 & 87.194 & 79.967 \\
		   & (7.55) & (1.07) & (1.11) & (8.56) & (2.14) & (2.23) \\
		10 & 0.019 & 241.525 & 477.246 & 0.036 & 490.085 & 958.478 \\
		   & (1.14) & (1.26) & (1.22) & (2.14) & (2.55) & (2.45) \\
		11 & 0.001 & 111.166 & 206.044 & 0.001 & 228.869 & 408.883 \\
		   & (1.03) & (1.20) & (1.17) & (2.03) & (2.48) & (2.33) \\
	\hline
	\hline
		\end{tabular}%
		\vspace{0.5em}
		\begin{minipage}{\linewidth}
			\footnotesize \textit{Note:} This table reports the two naive monthly trading-volume aggregates, one from the \texttt{OrdersMatched} logs and one from the \texttt{OrderFilled} logs. For each month, the first row reports the naive volume in millions of USD, and the second row, in parentheses, reports its ratio to the corresponding turnover ($V^E$) in \autoref{tab:monthly_vol_flow}. For the \texttt{OrderFilled} aggregate this ratio is the naive-to-turnover wedge of \autoref{subsec:crossmarket}.
	\end{minipage}
\end{table}


\begin{table}[htbp!]
	\centering
	\caption{The Naive-to-Turnover Wedge Across 249 Election-Related Markets}
	\label{tab:crossmarket_dist}
	\setlength{\tabcolsep}{6pt}
	\renewcommand{\arraystretch}{1.1}
	\begin{tabular}{lccccc}
		\hline
		\hline
		\multicolumn{6}{l}{\textit{Panel A: Wedge distribution by activity threshold}} \\
		\midrule
		Sample & $N$ & p10 & Median & p90 & Max \\
		\midrule
		All markets                      & 249 & 2.39 & 3.57 & 8.44 & 50.2 \\
		Lifetime naive $\geq$ \$0.1M     & 130 & 2.21 & 3.08 & 4.85 & 50.2 \\
		Lifetime naive $\geq$ \$1M       & 73  & 2.19 & 2.94 & 3.79 & 5.18 \\
		\midrule
		\multicolumn{6}{l}{\textit{Panel B: By size tercile (markets with naive $\geq$ \$0.1M)}} \\
		\midrule
		 & $N$ & Median wedge & Median $\kappa$ & & \\
		\midrule
		Small tercile  & 43 & 3.40 & 1.40 & & \\
		Middle tercile & 43 & 3.16 & 1.16 & & \\
		Large tercile  & 44 & 2.84 & 0.84 & & \\
		\hline
		\hline
	\end{tabular}
	\vspace{0.5em}
	\begin{minipage}{\linewidth}
		\footnotesize \textit{Note:} This table applies the decomposition to all 249 markets linked to the 2024 presidential election (12.9 million \texttt{OrderFilled} logs), spanning candidate, popular-vote, electoral-margin, state-winner, nomination, and exit-timing events. The wedge is lifetime naive volume over lifetime exchange-equivalent turnover, which by the identity in \autoref{subsec:market_measures} equals $2 + \kappa$. Terciles are formed on lifetime naive volume.
	\end{minipage}
\end{table}


\begin{table}[htbp!]
	\centering
	\caption{The Decomposition Applied in a Single-Market Event}
	\label{tab:binary_market}
	\setlength{\tabcolsep}{8pt}
	\renewcommand{\arraystretch}{1.1}
	\begin{tabular}{lc@{\hspace{3em}}lc}
		\hline
		\hline
		\multicolumn{2}{l}{\textit{Route components (\$M)}} & \multicolumn{2}{l}{\textit{Transaction types (count)}} \\
		\midrule
		Exchange component & 1.573 & Simple exchange & 2,958 \\
		Mint volume     & 4.496 & Pure minting    & 2,261 \\
		Burn volume     & 1.115 & Mixed with minting & 632 \\
		                &       & Pure burning    & 553 \\
		                &       & Mixed with burning & 251 \\
		\midrule
		\multicolumn{4}{l}{\textit{Aggregate measures (\$M)}} \\
		\midrule
		Naive \texttt{OrderFilled} volume & 8.757 & Naive \texttt{OrdersMatched} volume & 4.355 \\
		Exchange-equivalent turnover $V^E$ & 2.688 & Net inflow $F$ & 3.380 \\
		Market activity $V^G$ & 6.069 & Issuance-to-turnover $\kappa$ & 1.26 \\
		\hline
		\hline
	\end{tabular}
	\vspace{0.5em}
	\begin{minipage}{\linewidth}
		\footnotesize \textit{Note:} This table applies the transaction-level decomposition to the sole market in the single-market event ``Will Biden announce resignation by July 31?'' The event uses the base \texttt{CTFExchange} settlement format, with no negative-risk adapter and no conversion function. Its market contains 16,876 \texttt{OrderFilled} logs. For the transaction-type counts, total buys and sells that differ by no more than one raw unit ($10^{-6}$ USDC) are treated as equal. An issuance transaction is pure when its exchange component is zero and mixed when that component is positive.
	\end{minipage}
\end{table}


\begin{table}[htbp!]
	\centering
	\caption{What Predicts Issuance Intensity $\kappa = (V^G - V^E)/V^E$}
	\label{tab:crossmarket_reg}
	\setlength{\tabcolsep}{8pt}
	\renewcommand{\arraystretch}{1.1}
	\begin{tabular}{lcccc}
		\hline
		\hline
		Dependent variable: $\ln \kappa$ & \multicolumn{2}{c}{Market cross-section} & \multicolumn{2}{c}{Market-month panel} \\
		\cmidrule(lr){2-3} \cmidrule(lr){4-5}
		 & (1) & (2) & (3) & (4) \\
		\midrule
		$\ln(\text{naive volume})$ & $-0.214^{***}$ & $-0.087^{*}$   & $-0.128^{***}$ & $-0.111$ \\
		                           & (0.047)        & (0.051)        & (0.035)        & (0.076)  \\
		$\ln(\text{lifespan, days})$ &              & $-0.283^{***}$ &                &          \\
		                           &                & (0.099)        &                &          \\
		Single-market event          &                & $-0.724^{**}$  &                &          \\
		                           &                & (0.338)        &                &          \\
		$\ln(\text{event size})$   &                & $-0.080$       &                &          \\
		                           &                & (0.115)        &                &          \\
		Market age (months)        &                &                & $-0.123^{***}$ & $-0.094^{*}$ \\
		                           &                &                & (0.025)        & (0.055)  \\
		\midrule
		Market fixed effects       & no             & no             & no             & yes      \\
		$R^2$                      & 0.145          & 0.226          & 0.138          & 0.365    \\
		Observations               & 130            & 130            & 825            & 825      \\
		\hline
		\hline
	\end{tabular}
	\vspace{0.5em}
	\begin{minipage}{\linewidth}
		\footnotesize \textit{Note:} Columns (1) and (2) regress the log of lifetime issuance intensity on market characteristics across the 130 markets with lifetime naive volume of at least \$0.1 million, with heteroskedasticity-robust standard errors. The single-market-event indicator equals one for the sole market in its event. Event size is the number of markets within the same event. Columns (3) and (4) use the market-month panel (market-months with positive turnover, nonzero net inflow, and naive volume of at least \$0.01 million), with market age measured in months since the market's first trade and standard errors clustered by market. $^{***}$, $^{**}$, and $^{*}$ denote significance at the 1, 5, and 10 percent levels.
	\end{minipage}
\end{table}

\begin{table}[htbp!]
	\centering
	\caption{Monthly Kyle's Lambda under Naive and Corrected Order Flow}
	\label{tab:lambda_monthly}
	\setlength{\tabcolsep}{5pt}
	\renewcommand{\arraystretch}{1.1}
	\begin{tabular}{lcccc}
		\hline
		\hline
		 & \multicolumn{2}{c}{$\hat\lambda$ per \$1M} & \multicolumn{2}{c}{Cost of moving 0.50 to 0.55 (\$M)} \\
		\cmidrule(lr){2-3} \cmidrule(lr){4-5}
		Month (2024) & Naive & Corrected & Naive & Corrected \\
		\midrule
		January   & 0.539 (0.270) & 0.587 (0.312) & 0.372 & 0.342 \\
		February  & 0.205 (0.034) & 0.439 (0.072) & 0.978 & 0.457 \\
		March     & 0.315 (0.086) & 0.489 (0.121) & 0.637 & 0.410 \\
		April     & 0.085 (0.022) & 0.153 (0.055) & 2.358 & 1.314 \\
		May       & 0.241 (0.048) & 0.352 (0.075) & 0.834 & 0.571 \\
		June      & 0.278 (0.176) & 0.384 (0.241) & 0.722 & 0.523 \\
		July      & 0.243 (0.093) & 0.421 (0.120) & 0.824 & 0.476 \\
		August    & 0.076 (0.012) & 0.132 (0.016) & 2.646 & 1.520 \\
		September & 0.038 (0.009) & 0.048 (0.020) & 5.246 & 4.170 \\
		October   & 0.013 (0.004) & 0.022 (0.005) & 15.629 & 9.116 \\
		\hline
		\hline
	\end{tabular}
	\vspace{0.5em}
	\begin{minipage}{\linewidth}
		\footnotesize \textit{Note:} This table reports calendar-month estimates of Kyle's $\lambda$ for the Trump YES share from regressions of 15-minute log-odds price changes on 15-minute net signed order flow in million USD, with Newey-West standard errors in parentheses. The Naive column signs the raw \texttt{OrderFilled} aggregate, and the Corrected column signs the one-count flow of exchange plus mint plus burn. The cost columns give the net purchase required to move the price from 0.50 to 0.55 at the estimated $\lambda$.
	\end{minipage}
\end{table}


\begin{table}[htbp!]
	\centering
	\caption{Secondary versus Primary Price Impact by Market Activity}
	\label{tab:price_layers}
	\setlength{\tabcolsep}{4pt}
	\renewcommand{\arraystretch}{1.1}
	\begin{tabular}{lccccc}
		\hline
		\hline
		 & Q1 (thinnest) & Q2 & Q3 & Q4 & Q5 (deepest) \\
		\midrule
		Median $V^G$ (\$M) & 0.02 & 0.04 & 0.10 & 0.96 & 6.49 \\
		\addlinespace
		Secondary $\lambda$ (exchange) & $630.6^{**}$ & $445.0^{***}$ & $156.1^{***}$ & $15.6^{***}$ & $0.19^{**}$ \\
		 & (307.2) & (169.4) & (36.3) & (5.0) & (0.09) \\
		Primary $\lambda$ (issuance) & $374.5^{***}$ & $196.5^{**}$ & $37.3^{**}$ & $6.3$ & $0.08$ \\
		 & (97.7) & (79.5) & (15.5) & (4.0) & (0.05) \\
		\addlinespace
		Difference $p$-value & 0.46 & 0.22 & $<$0.01 & 0.06 & 0.07 \\
		Markets & 43 & 42 & 43 & 42 & 43 \\
		Observations & 37{,}414 & 38{,}241 & 269{,}647 & 1{,}037{,}801 & 1{,}053{,}665 \\
		\hline
		\hline
	\end{tabular}
	\vspace{0.5em}
	\begin{minipage}{\linewidth}
		\footnotesize \textit{Note:} The negative-risk election markets with an estimable price impact are sorted into quintiles by lifetime activity $V^G$, from the thinnest (Q1) to the deepest (Q5). Within each quintile, 15-minute log-odds price changes are regressed on the secondary (exchange trade) and primary (issuance) components of signed order flow, with market fixed effects. Both $\lambda$ rows are in log odds per \$1 million, with standard errors clustered by market in parentheses. The difference row reports the $p$-value of the test that the two coefficients are equal. $^{***}$, $^{**}$, and $^{*}$ denote significance at the 1, 5, and 10 percent levels.
	\end{minipage}
\end{table}


\begin{table}[htbp!]
	\centering
	\caption{Monthly Participation in the Three Candidate Markets}
	\label{tab:monthly_participation}
	\setlength{\tabcolsep}{8pt}
	\renewcommand{\arraystretch}{1.1}
	\resizebox{\textwidth}{!}{%
	\begin{tabular}{ccccc}
	\hline
	\hline
	Month & Active addresses & First-time addresses & Market activity ($V^G$, \$M) & $V^G$ per address (USD) \\
	\midrule
	1  & 460     & 460     & 4.125   & 8,968  \\
	2  & 588     & 454     & 2.737   & 4,655  \\
	3  & 526     & 359     & 3.048   & 5,794  \\
	4  & 394     & 231     & 3.970   & 10,076 \\
	5  & 3,388   & 3,139   & 4.917   & 1,451  \\
	6  & 6,181   & 4,942   & 8.232   & 1,332  \\
	7  & 14,622  & 11,459  & 40.680  & 2,782  \\
	8  & 23,314  & 18,337  & 54.040  & 2,318  \\
	9  & 33,087  & 23,389  & 90.475  & 2,734  \\
	10 & 139,064 & 117,234 & 865.624 & 6,225  \\
	11 & 94,307  & 51,881  & 369.823 & 3,921  \\
	\hline
	\hline
	\end{tabular}%
	}
	\vspace{0.5em}
	\begin{minipage}{\linewidth}
		\footnotesize \textit{Note:} This table counts the addresses that traded the Trump, Biden, or Harris market in each month of the sample. An address is active in a month if it is the maker of at least one order filled that month. An address is first-time in the month in which it first appears in these three markets. The market activity column sums the three markets' $V^G$ from \autoref{tab:monthly_vol_flow}, and the last column divides it by the month's active addresses. The sample ends on November~6, 2024, so month~11 contains fewer than six trading days.
	\end{minipage}
\end{table}


\begin{table}[htbp!]
	\centering
	\caption{The October Whale and Buyer Concentration}
	\label{tab:whale_concentration}
	\setlength{\tabcolsep}{6pt}
	\renewcommand{\arraystretch}{1.1}
	\begin{tabular}{lccc}
		\hline
		\hline
		\multicolumn{4}{l}{\textit{Panel A: The four whale accounts (October Trump YES net buying)}} \\
		\midrule
		Account & Net buy (\$M) & Net buy / $F$ & Net buy / $V^G$ \\
		\midrule
		Fredi9999    & 10.432 & 5.913\% & 1.838\% \\
		Theo4        & 7.911 & 4.484\% & 1.394\% \\
		PrincessCaro & 5.030 & 2.851\% & 0.887\% \\
		Michie       & 3.259 & 1.848\% & 0.574\% \\
		\textbf{Whale total} & \textbf{26.633} & \textbf{15.096\%} & \textbf{4.693\%} \\
		\midrule
		\multicolumn{4}{l}{\textit{Panel B: Concentration of October net buying (non-market-maker accounts)}} \\
		\midrule
		 & Net buy (\$M) & Net buy / $F$ & Net buy / $V^G$ \\
		\midrule
		Top 1   & 16.998 & 9.635\% & 2.995\% \\
		Top 5   & 54.843 & 31.087\% & 9.665\% \\
		Top 10  & 71.642 & 40.609\% & 12.625\% \\
		Top 20  & 89.405 & 50.678\% & 15.756\% \\
		\hline
		\hline
	\end{tabular}
	\vspace{0.5em}
	\begin{minipage}{\linewidth}
		\footnotesize \textit{Note:} Panel A reports October 2024 net purchases of Trump YES exposure for the four accounts of the French trader reported by the press. Panel B reports the concentration of October net buying among non-market-maker accounts, after excluding the 109 behavioral market-making accounts. In both panels, the ratio columns report net buys as a share of October Trump-market net inflow $F$ and market activity $V^G$.
	\end{minipage}
\end{table}

\clearpage

\appendix
\counterwithin{figure}{section}
\counterwithin{table}{section}
\setcounter{page}{1}

\begin{center}
{\Large\bfseries Online Appendix}
\end{center}

\noindent \autoref{appendix_protocol} documents Polymarket's protocol mechanics, the order-flow steps and the split, merge, and conversion operations. \autoref{appendix_tx_decomposition} states the algorithm that classifies each transaction and computes its exchange, minting, and burning components. \autoref{appendix_robustness} reports the window sensitivity of exchange-equivalent turnover and the vector-autoregression check on the price-impact estimates. \autoref{appendix_uniqueness} shows that exchange-equivalent turnover is the only window measure that counts each transferred dollar once, is unaffected by the engine's routing, and assigns no turnover to unoffset issuance.

\section{Protocol Details}
\label{appendix_protocol}

This appendix documents the Polymarket protocol mechanics summarized in \autoref{polymarket}.

\subsection{Market structure and order flow}

\autoref{polymarket} defines the event and market hierarchy and the two settlement formats. The 2024 Presidential Election Winner event contains 17 markets: Donald Trump, Joe Biden, Nikki Haley, Gavin Newsom, Robert F. Kennedy Jr., Ron DeSantis, Kamala Harris, Vivek Ramaswamy, Michelle Obama, Hillary Clinton, Chris Christie, AOC, Elizabeth Warren, Bernie Sanders, Other Democrat Politician, Other Republican Politician, and Kanye.

Order flow proceeds in five steps. (1) A trader signs an EIP-712-compliant order message off-chain, specifying the asset, amount, and price. (2) The Polymarket operator places the signed message in the off-chain central limit order book. (3) The matching engine finds one or more complementary orders. (4) The operator submits the signed orders to the appropriate exchange contract, \texttt{CTFExchange} for markets in single-market events or \texttt{NegRiskCTFExchange} for markets in multi-market events, in a single on-chain transaction. (5) The contract atomically executes the match and emits \texttt{OrderFilled} logs for each order plus an \texttt{OrdersMatched} summary log. Outcome shares are ERC-1155 conditional tokens created under the Gnosis Conditional Token Framework (\texttt{CTF}). Upon resolution, winning shares redeem for 1 USDC each from locked collateral.

\subsection{Share minting: position split}

Share minting and share burning, the two issuance routes defined in \autoref{polymarket}, are carried out by the \texttt{splitPosition(.)} and \texttt{mergePositions(.)} functions of the Gnosis \texttt{CTF} contract. The \texttt{splitPosition(.)} function creates new outcome shares, recorded as \texttt{PositionSplit} events, and operates in two ways.

(1) Counterparty order matching. Suppose Trader X is willing to pay 0.70 USDC for a YES share of candidate A and Trader Y is willing to pay 0.30 USDC for a NO share of candidate A. After matching, the exchange contract calls \texttt{splitPosition(.)},\footnote{In multi-market events, share minting and merging run through the \texttt{NegRiskAdapter} contract, which records them as \texttt{PositionSplit} and \texttt{PositionsMerge} logs.} the combined 1 USDC is locked as collateral, and a new (1 YES, 1 NO) pair is minted and distributed to the two traders. Matching complementary buys through minting is the main method of share creation and underlies the market's ``\$1.00 Rule.''

(2) Unilateral arbitrage. If the combined market price of a YES-NO pair exceeds \$1.00, an arbitrageur can call \texttt{splitPosition(.)} directly, deposit 1 USDC, receive the pair, and sell both legs, pressing prices down.

\subsection{Share burning: position merge}

The \texttt{mergePositions(.)} function is the inverse operation, recorded as \texttt{PositionsMerge} events. (1) Counterparty order matching: two traders selling opposing shares are matched, the complete pair is burned, and the 1 USDC collateral is released and split at the traded prices. (2) Unilateral arbitrage: if the pair trades below \$1.00, an arbitrageur buys both legs and merges them for 1 USDC, pressing prices up. Together, split and merge keep $P^{\text{YES}} + P^{\text{NO}}$ near \$1.00.

\subsection{Position conversion}

Multi-market events require an additional mechanism to keep prices consistent across markets, so that YES prices sum to \$1 across the event. The \texttt{convertPositions(.)} function in the \texttt{NegRiskAdapter} contract, recorded as \texttt{PositionsConverted} events, lets a trader transform a portfolio of NO shares into the economically equivalent portfolio of YES shares plus USDC. \autoref{tab_economic_equivalence_portfolios} illustrates the equivalence for a three-outcome event: 1 NO A plus 1 NO B pays the same as 1 YES C plus 1 USDC in every state. More generally, in an event with $N$ outcomes, converting $Q$ units of NO shares across $M$ outcomes yields $Q$ units of YES shares for each of the remaining $N-M$ outcomes plus $(M-1)\times Q$ USDC. Conversion improves capital efficiency and enables arbitrage across markets that disciplines relative prices.

\begin{table}[!htbp]
	\centering
	\caption{Economic Equivalence of Portfolios in a Three-Outcome Event}
	\label{tab_economic_equivalence_portfolios}
	\begin{tabular}{l|c|c}
		\hline
		\hline
		\text{Resolution} & \text{Portfolio 1: } & \text{Portfolio 2: } \\
		                                  & \text{1 NO A + 1 NO B} & \text{1 YES C + 1 USDC} \\
		\hline
		\text{A Wins} & \$0 \text{ (NO A)} + \$1 \text{ (NO B)} = \text{\$1} & \$0 \text{ (YES C)} + \$1 \text{ (USDC)} = \text{\$1} \\
		\text{B Wins} & \$1 \text{ (NO A)} + \$0 \text{ (NO B)} = \text{\$1} & \$0 \text{ (YES C)} + \$1 \text{ (USDC)} = \text{\$1} \\
		\text{C Wins} & \$1 \text{ (NO A)} + \$1 \text{ (NO B)} = \text{\$2} & \$1 \text{ (YES C)} + \$1 \text{ (USDC)} = \text{\$2} \\
		\hline
		\hline
	\end{tabular}
	\vspace{0.5em}
	\begin{minipage}{\linewidth}
		\footnotesize \textit{Note:} This table shows the payoff at resolution for two economically equivalent portfolios in a three-outcome prediction event (candidates A, B, and C), illustrating the logic behind the \texttt{convertPositions(.)} function. Both portfolios yield identical payoffs under all possible outcomes.
	\end{minipage}
\end{table}

\section{Transaction-Level Decomposition Algorithm}
\label{appendix_tx_decomposition}

This appendix states the algorithm that implements the transaction-level decomposition of \autoref{sec:volume_measurement}.

\begin{algorithm}[htbp!]
	\caption{Transaction-Level Volume Decomposition}
	\begin{spacing}{1.25}
	\small
	\begin{algorithmic}
	\Require Fill records $\mathcal{F}=\{(m_i,t_i,M_i,T_i)\}_{i=1}^{K}$ from one transaction; outcome share IDs $y$ (YES) and $n$ (NO), with USDC ID $0$
	\Ensure $V^{\mathrm{tr}}(y),V^{\mathrm{tr}}(n),V^{\mathrm{mint}}(y),V^{\mathrm{mint}}(n),V^{\mathrm{burn}}(y),V^{\mathrm{burn}}(n)$

	\State $\mathcal{I}(a)\gets\{i:m_i=0,\;t_i=a\}$ and $\mathcal{J}(a)\gets\{i:t_i=0,\;m_i=a\}$ for $a\in\{y,n\}$
	\State $B(a)\gets 10^{-6}\sum_{i\in\mathcal{I}(a)} M_i$ and $S(a)\gets 10^{-6}\sum_{i\in\mathcal{J}(a)} T_i$ for $a\in\{y,n\}$
	\State $V^{\mathrm{buy}}\gets B(y)+B(n)$;\;\; $V^{\mathrm{sell}}\gets S(y)+S(n)$;\;\; $v^{\mathrm{tr}}\gets \min\{V^{\mathrm{buy}},V^{\mathrm{sell}}\}$
	\State Initialize $V^{\mathrm{tr}}(a),V^{\mathrm{mint}}(a),V^{\mathrm{burn}}(a)\gets 0$ for $a\in\{y,n\}$
		\State Let $a^{\mathrm{buy}}\in\{t_i \mid m_i = 0\}$ and $a^{\mathrm{sell}}\in\{m_i \mid t_i = 0\}$ be arbitrary, each defined only when its set is nonempty; each branch below uses only the selector that is defined.

		\If{$V^{\mathrm{buy}} > V^{\mathrm{sell}}$} \Comment{Share minting}
			\State $a^{\star}\gets a^{\mathrm{sell}}$ if $\bigl|\{m_i : i=1,\dots,K\}\bigr| > 1$, else $a^{\mathrm{buy}}$;\;\; $\bar a\gets y$ if $a^{\star}=n$, else $n$
		\State $V^{\mathrm{tr}}(a^{\star})\gets v^{\mathrm{tr}}$
		\State $V^{\mathrm{mint}}(a^{\star})\gets B(a^{\star})-v^{\mathrm{tr}}$;\;\; $V^{\mathrm{mint}}(\bar a)\gets B(\bar a)$
		\ElsIf{$V^{\mathrm{buy}} < V^{\mathrm{sell}}$} \Comment{Share burning}
			\State $a^{\star}\gets a^{\mathrm{buy}}$ if $\bigl|\{t_i : i=1,\dots,K\}\bigr| > 1$, else $a^{\mathrm{sell}}$;\;\; $\bar a\gets y$ if $a^{\star}=n$, else $n$
		\State $V^{\mathrm{tr}}(a^{\star})\gets v^{\mathrm{tr}}$
		\State $V^{\mathrm{burn}}(a^{\star})\gets S(a^{\star})-v^{\mathrm{tr}}$;\;\; $V^{\mathrm{burn}}(\bar a)\gets S(\bar a)$
	\Else \Comment{Pure exchange}
		\State $a^{\star}\gets a^{\mathrm{buy}}$;\;\; $V^{\mathrm{tr}}(a^{\star})\gets v^{\mathrm{tr}}$
	\EndIf

	\State \Return $V^{\mathrm{tr}}(y),V^{\mathrm{tr}}(n),V^{\mathrm{mint}}(y),V^{\mathrm{mint}}(n),V^{\mathrm{burn}}(y),V^{\mathrm{burn}}(n)$
	\end{algorithmic}
	\normalsize
	\end{spacing}
\end{algorithm}

\section{Robustness}
\label{appendix_robustness}

\subsection{Window sensitivity of exchange-equivalent turnover}

Because $V^E$ nets mints against burns within a window, its level could in principle depend on how long that window is. \autoref{tab:window_sensitivity} reports October 2024 Trump-market $V^E$ under alternative netting windows.

\subsection{Order-flow feedback and the dynamic price response}
\label{appendix_var}

The estimating equation of \autoref{subsec:lambda_estimation} relates an interval's price change to the same interval's net order flow. If flow itself responds to past price moves, that static regression could fold feedback trading into $\hat\lambda$. We therefore estimate, month by month, a bivariate vector autoregression in flow and returns with flow ordered first,
\begin{equation}
	\begin{bmatrix} Q_t \\ \Delta\theta_t \end{bmatrix}
	= c + \sum_{k=1}^{48} A_k \begin{bmatrix} Q_{t-k} \\ \Delta\theta_{t-k} \end{bmatrix} + u_t,
	\qquad \mathrm{E}\left[u_t u_t'\right] = \Sigma,
\end{equation}
using 48 lags, twelve hours of 15-minute intervals. Ordering flow first lets a flow innovation move the price within the same interval, which is the response the static $\lambda$ measures. Let $P$ be the Cholesky factor of $\Sigma$ and $\Phi_h$ the moving-average coefficient matrices of the system. The cumulative price response to a flow innovation over 96 intervals, one trading day, is
\begin{equation}
	\lambda^{\mathrm{VAR}} = \frac{1}{P_{11}} \sum_{h=0}^{96} \left[\Phi_h P\right]_{21},
\end{equation}
where dividing by $P_{11}$, the standard deviation of the flow innovation, rescales the impulse from one standard deviation to \$1 million so that the estimate is comparable with $\hat\lambda$.

\autoref{tab:lambda_var} sets the two against each other for the corrected flow. The dynamic estimates are of the same order as the static ones in every month and larger in seven of the ten, so allowing flow to respond to past prices does not systematically shrink the measured price impact. In October, the month behind the paper's cost figures, the two agree within 5 percent.

\begin{table}[!htbp]
	\centering
	\caption{October 2024 Trump-Market $V^E$ under Alternative Netting Windows}
	\label{tab:window_sensitivity}
	\begin{tabular}{lc}
		\hline
		\hline
		Netting window & $V^E$ (\$M) \\
		\midrule
		None (pure exchange component only) & 268.4 \\
		Daily   & 391.0 \\
		Weekly  & 391.0 \\
		Monthly & 391.0 \\
		\hline
		\hline
	\end{tabular}
	\vspace{0.5em}
	\begin{minipage}{\linewidth}
		\footnotesize \textit{Note:} Each row computes $V^E$ by netting aggregate mints against aggregate burns within windows of the stated length and summing over October 2024, so the rows show its sensitivity to the netting horizon. The first row reports the exchange component alone.
	\end{minipage}
\end{table}


\begin{table}[!htbp]
	\centering
	\caption{Static and Dynamic Price Impact under Corrected Order Flow}
	\label{tab:lambda_var}
	\renewcommand{\arraystretch}{1.1}
	\begin{tabular}{lcc}
		\hline
		\hline
		Month (2024) & Static $\hat\lambda$ & $\lambda^{\mathrm{VAR}}$ \\
		\midrule
		January   & 0.587 & 0.785 \\
		February  & 0.439 & 0.234 \\
		March     & 0.489 & 0.485 \\
		April     & 0.153 & 0.322 \\
		May       & 0.352 & 0.532 \\
		June      & 0.384 & 0.244 \\
		July      & 0.421 & 0.498 \\
		August    & 0.132 & 0.208 \\
		September & 0.048 & 0.057 \\
		October   & 0.022 & 0.023 \\
		\hline
		\hline
	\end{tabular}
	\vspace{0.5em}
	\begin{minipage}{\linewidth}
		\footnotesize \textit{Note:} Both columns are in log odds per million USD of corrected net order flow. The static column repeats the Corrected column of \autoref{tab:lambda_monthly}. The $\lambda^{\mathrm{VAR}}$ column reports the cumulative 24-hour price response to a \$1 million innovation in corrected flow, from the bivariate vector autoregression above with 48 lags and the flow innovation ordered first. November is omitted as in \autoref{tab:lambda_monthly}.
	\end{minipage}
\end{table}

\clearpage
\section{Properties of Exchange-Equivalent Turnover}
\label{appendix_uniqueness}

This appendix states three properties a window turnover measure should have and checks that exchange-equivalent turnover (\autoref{subsec:market_measures}) is the only measure with all three.

Fix one share type of one market and a measurement window. Aggregating the transaction-level decomposition over the window yields three route totals, exchange dollars $x$, minting dollars $m$, and burning dollars $b$, each measured in the traded leg's USDC value. A window turnover measure is a function $V(x, m, b)$. Consider three properties.

(1) \emph{Single counting.} In a window with no issuance activity, turnover is the dollars exchanged: $V(x, 0, 0) = x$.

(2) \emph{Routing invariance.} The measure is unchanged when an exchange of $s$ dollars is rerouted through an offsetting mint and burn of $s$ dollars within the window, or the reverse: $V(x - s, m + s, b + s) = V(x, m, b)$. The matching engine, not the trader, chooses the route (\autoref{subsec:market_measures}), so a turnover measure should not depend on that choice.

(3) \emph{Residual issuance is not turnover.} Issuance that is not offset within the window transfers no existing claim between traders, so it adds nothing: $V(x, m, 0) = V(x, 0, b) = V(x, 0, 0)$.

These three properties leave a single measure. Write $\mu = \min(m, b)$. Property (2) reroutes the offsetting issuance back into exchange, so $V(x, m, b) = V(x + \mu, m - \mu, b - \mu)$, where at least one issuance total is now zero. Property (3) then drops the remaining issuance and property (1) evaluates the rest, giving $V(x, m, b) = x + \mu$. This is the exchange-equivalent turnover of \autoref{subsec:market_measures}, $V^E = x + \min(m, b)$, taken leg by leg and summed over a market's two share types. The derivation fixes the value at every $(x, m, b)$, so no other measure has all three properties.

Each alternative aggregate fails one of the properties. The naive \texttt{OrderFilled} aggregate $2x + m + b$ counts exchanged dollars twice and fails single counting. Market activity $V^G = x + \max(m, b)$ counts residual issuance and fails property (3). The \texttt{OrdersMatched} aggregate depends on which side of each trade crossed the spread and therefore on the route, and fails routing invariance. Property (2) also has a direct economic reading. A mint and an offsetting burn within the window are the settlement contract intermediating a transfer from the burn-side seller to the mint-side buyer, standing where a dealer's inventory would stand in a conventional market, and the measure counts that intermediated transfer once, as it would count a direct one.

\clearpage

\end{document}